\documentclass[a4paper,prb,amsmath,amssymb]{revtex4}
\usepackage[english]{babel}
\usepackage[latin1]{inputenc}
\usepackage{amscd,amsmath,amssymb,verbatim}
\usepackage[dvips]{graphicx,color}
\usepackage{subfigure}
\usepackage{textcomp}
\usepackage{calc}
\usepackage{bbm}
\usepackage[OT1,T1]{fontenc}
\usepackage{natbib}
\usepackage[verbose,hypertexnames=false,bookmarksopenlevel=1,filecolor=blue,
linkcolor=blue,citecolor=blue,pdfstartview=FitH,bookmarksopen,bookmarksnumbered,
colorlinks,plainpages=false,linktocpage,ps2pdf]{hyperref}

\DeclareMathAlphabet{\mathpzc}{OT1}{pzc}{m}{it}

\DeclareSymbolFont{mathbold}{OML}{cmm}{b}{it}
\DeclareMathSymbol{\balpha}{\mathord}{mathbold}{11}
\DeclareMathSymbol{\bbeta}{\mathord}{mathbold}{12}
\DeclareMathSymbol{\bgamma}{\mathord}{mathbold}{13}
\DeclareMathSymbol{\bdelta}{\mathord}{mathbold}{14}
\DeclareMathSymbol{\bepsilon}{\mathord}{mathbold}{15}
\DeclareMathSymbol{\bvarepsilon}{\mathord}{mathbold}{34}
\DeclareMathSymbol{\bzeta}{\mathord}{mathbold}{16}
\DeclareMathSymbol{\bEta}{\mathord}{mathbold}{17}
\DeclareMathSymbol{\btheta}{\mathord}{mathbold}{18}
\DeclareMathSymbol{\bvartheta}{\mathord}{mathbold}{35}
\DeclareMathSymbol{\biota}{\mathord}{mathbold}{19}
\DeclareMathSymbol{\bkappa}{\mathord}{mathbold}{20}
\DeclareMathSymbol{\blambda}{\mathord}{mathbold}{21}
\DeclareMathSymbol{\bmu}{\mathord}{mathbold}{22}
\DeclareMathSymbol{\bnu}{\mathord}{mathbold}{23}
\DeclareMathSymbol{\bxi}{\mathord}{mathbold}{24}
\DeclareMathSymbol{\bpi}{\mathord}{mathbold}{25}
\DeclareMathSymbol{\bvarpi}{\mathord}{mathbold}{36}
\DeclareMathSymbol{\brho}{\mathord}{mathbold}{26}
\DeclareMathSymbol{\bvarrho}{\mathord}{mathbold}{37}
\DeclareMathSymbol{\bsigma}{\mathord}{mathbold}{27}
\DeclareMathSymbol{\bvarsigma}{\mathord}{mathbold}{38}
\DeclareMathSymbol{\btau}{\mathord}{mathbold}{28}
\DeclareMathSymbol{\bupsilon}{\mathord}{mathbold}{29}
\DeclareMathSymbol{\bphi}{\mathord}{mathbold}{30}
\DeclareMathSymbol{\bvarphi}{\mathord}{mathbold}{39}
\DeclareMathSymbol{\bchi}{\mathord}{mathbold}{31}
\DeclareMathSymbol{\bpsi}{\mathord}{mathbold}{32}
\DeclareMathSymbol{\bomega}{\mathord}{mathbold}{33}
\DeclareMathSymbol{\biGamma}{\mathord}{mathbold}{0}
\DeclareMathSymbol{\biDelta}{\mathord}{mathbold}{1}
\DeclareMathSymbol{\biTheta}{\mathord}{mathbold}{2}
\DeclareMathSymbol{\biLambda}{\mathord}{mathbold}{3}
\DeclareMathSymbol{\biXi}{\mathord}{mathbold}{4}
\DeclareMathSymbol{\biPi}{\mathord}{mathbold}{5}
\DeclareMathSymbol{\biSigma}{\mathord}{mathbold}{6}
\DeclareMathSymbol{\biUpsilon}{\mathord}{mathbold}{7}
\DeclareMathSymbol{\biPhi}{\mathord}{mathbold}{8}
\DeclareMathSymbol{\biPsi}{\mathord}{mathbold}{9}
\DeclareMathSymbol{\biOmega}{\mathord}{mathbold}{10}

\newcommand{\ket}[1]{|#1\rangle}

\usepackage{graphicx}
\usepackage{epsfig}

\begin{document}
\title{ Spin Relaxation  in Quantum Wires  }

\author{P. Wenk}
\email[]{p.wenk@jacobs-university.de}
\homepage[]{www.physnet.uni-hamburg.de/hp/pwenk/}
\affiliation{School of Engineering and Science, Jacobs University Bremen, Bremen 28759, Germany}
\author{S. Kettemann}
\email[]{s.kettemann@jacobs-university.de}
\homepage[]{www.jacobs-university.de/ses/skettemann}
\affiliation{School of Engineering and Science, Jacobs University Bremen, Bremen 28759, Germany, and Asia Pacific Center for Theoretical
Physics and Division of Advanced Materials Science Pohang University of Science and Technology (POSTECH) San31, Hyoja-dong, Nam-gu, Pohang 790-784, South Korea}

\begin{abstract} 
The spin dynamics and spin relaxation  of itinerant  electrons in  quantum wires with spin-orbit coupling is reviewed.  We give  an introduction to spin dynamics, and review  spin-orbit coupling mechanisms in semiconductors. The spin diffusion equation with spin-orbit coupling is derived, using only intuitive, classical  random walk arguments. We give an overview of all spin relaxation mechanisms, with particular emphasis on the motional narrowing mechanism in disordered conductors, the D'yakonov-Perel'-Spin relaxation (DPS). Here, we discuss in particular, the existence of persistent spin helix solutions of the spin diffusion equation, with vanishing spin relaxation rates. We then, derive solutions of the spin diffusion equation in quantum wires, and show that there is an effective alignment of the spin-orbit field in wires whose width is smaller than the spin precession length $L_{\rm SO}$. We show that the resulting reduction in the spin relaxation rate results in a change in the sign of the  quantum corrections  to the conductivity. Finally, we present recent experimental results which confirm the decrease of the spin relaxation rate in wires whose width is smaller than $L_{\rm SO}$: the direct optical measurement of the spin relaxation rate, as well as transport measurements, which show a dimensional crossover from weak antilocalization to weak localization as the wire width is reduced. Open problems remain, in particular in narrower, ballistic wires, were optical and transport measurements seem to find opposite behavior of the spin relaxation rate: enhancement,  suppression, respectively. We conclude with a review of these and other open problems which still challenge the theoretical understanding and modeling of the experimental results.
\end{abstract}

\maketitle 
\newpage
\tableofcontents
\newpage
\section{Introduction}
The emerging technology of spintronics intends to use the manipulation of the spin degree of freedom of individual electrons for energy efficient  storage and transport of  information.\cite{zutic04} In contrast to classical electronics, which relies on the steering of charge carriers through semiconductors, spintronics uses the spin carried by  electrons, resembling tiny spinning tops. The difference to a classical top is that its angular momentum is quantized, it can only take  two discrete values, up or down. To control the spin of electrons, a detailed understanding of the interaction between the spin and orbital degrees of freedom of electrons and other mechanisms which do not conserve its spin, is necessary. These are typically  weak perturbations,  compared to the kinetic energy of conduction electrons, so that their spin relaxes slowly to the  advantage of  spintronic applications. The relaxation, or depolarization of the electron spin can occur due to the randomization of the electron momentum by  scattering from impurities, and dislocations in the material, and  due to scattering with elementary excitations  of the solid such as phonons and  other  electrons, when it  is transferred to the randomization of the electron spin due to the spin-orbit interaction. In addition, scattering from localized spins, such as nuclear spins and magnetic impurities are sources of electron spin relaxation. The electron  spin relaxation can be  reduced by constraining the electrons in low dimensional structures, quantum wells (confined in one direction, free in two dimensions), quantum wires ( confined in two directions, free in one direction), or quantum dots ( confined in all three directions). Although spin relaxation is typically smallest in quantum dots due to their discrete energy level spectrum, the necessity to transfer  the spin in spintronic devices, recently lead to intense research efforts to reduce the spin relaxation in quantum wires, where the energy spectrum is continuous. In the following we will review the theory of  spin dynamics and relaxation in quantum wires, and compare it with  recent experimental results. After a general introduction to spin dynamics in Section\,\ref{sec:SpinDyn}, we discuss all relevant spin relaxation mechanisms and how they depend on dimension, temperature, mobility, charge carrier density and magnetic field in Section\,\ref{sec:SpinRelaxMech}. In particular, we review recent results on spin relaxation in  semiconducting quantum wires, and its influence on the quantum corrections to their conductance in Section\,\ref{sec:SpinDynQW}. These weak localization corrections are thereby a very sensitive  measure of spin relaxation in quantum wires, in addition to optical methods as we review in Section\,\ref{sec:ExpResults}.  We set $\hbar =1$ in the following.
\section{Spin Dynamics}\label{sec:SpinDyn}  
Before we review the spin dynamics of conduction electrons and holes in semiconductors and metals, let us first reconsider the spin dynamics of a localized spin, as governed by the Bloch equations. 
       \subsection{ Dynamics of a Localized Spin}
A localized spin ${\bf \hat{s}}$, like a nuclear spin, or  the spin of a magnetic impurity in a solid,  precesses in an external magnetic field ${\bf B}$ due to the Zeeman interaction  with Hamiltonian $H_Z = - \gamma_g {\bf \hat{s} B }$, where $\gamma_g$ is the corresponding gyromagnetic ratio of the nuclear spin or magnetic impurity spin, respectively, which we will set equal to one, unless needed explicitly. This spin  dynamics is governed by  the Bloch equation of a localized spin, 
\begin{equation}
\partial_t {\bf \hat{s}} = \gamma_g {\bf \hat{s}} \times {\bf B}.
\end{equation}
	This equation is identical to the 
	Heisenberg equation  $   \partial_t {\bf \hat{s}} = - i  [{\bf \hat{s}} , H_Z ]$ for the quantum mechanical spin operator ${\bf \hat{s}}$ of an $S=1/2$-spin,  interacting with 
	the external magnetic field ${\bf B}$ due to  the Zeeman interaction with Hamiltonian 
	$H_Z $. 
	The  solution of the Bloch equation for a magnetic field pointing in 
	the z-direction is $\hat{s}_z(t) = \hat{s}_z(0)$, while the x- and y- components of the 
	spin are precessing with frequency ${\bomega}_0 = \gamma_g {\bf B}$ around the z-axis,
	$\hat{s}_x(t) =\hat{s}_x(0) \cos \omega_0 t + \hat{s}_y(0) \sin \omega_0 t $, $\hat{s}_y(t) =- \hat{s}_x(0) \sin \omega_0 t + \hat{s}_y(0) \cos \omega_0 t $.
	Since a localized spin interacts with its environment by  exchange interaction
and magnetic dipole interaction, 
	the precession will dephase after  a time $\tau_2$, and the 
	z-component of the spin relaxes to its equilibrium value $s_{z 0}$
	within a relaxation  time  $\tau_1$. This modifies the 
	Bloch equations to the phenomenological equations, 
	\begin{align}
	\partial_t \hat{s}_x ={}& \gamma_g  (\hat{s}_y  B_z - \hat{s}_z B_y) - \frac{1}{\tau_2} \hat{s}_x \nonumber \\
	\partial_t \hat{s}_y  ={}& \gamma_g  (\hat{s}_z B_x - \hat{s}_x B_z) - \frac{1}{\tau_2} \hat{s}_y \nonumber \\
	\partial_t \hat{s}_z  ={}& \gamma_g  (\hat{s}_x  B_y - \hat{s}_y B_x) - \frac{1}{\tau_1} (\hat{s}_z- s_{z 0}).
	\end{align}

\subsection{Spin Dynamics of Itinerant  Electrons}
\subsubsection{Ballistic Spin Dynamics }
The intrinsic degree of freedom spin is a direct consequence of the Lorentz invariant formulation of quantum mechanics. Expanding   the relativistic Dirac equation in the ratio  of the electron velocity and the   constant velocity of light $c$, one obtains in addition to the Zeeman term, a term which couples the spin ${\bf s}$ with the momentum ${\bf p}$ of the electrons, the spin-orbit coupling 
               \begin{equation} \label{so}
               H_{\rm SO} = - \frac{\mu_B}{2m c^2} {\bf \hat{s}} ~  {\bf p} \times {\bf E} = -  {\bf \hat{s}} {\bf B}_{\rm SO} ({\bf p}) ,
               \end{equation} 
                where we set the gyromagnetic ratio $\gamma_g =1$.
                ${\bf E} = - {\bf \nabla} V $, is an electrical field, 
                 and ${\bf B}_{\rm SO} ({\bf p}) = \mu_B/(2 m c^2)  {\bf p} \times {\bf E} $.
                Substitution into the Heisenberg equation yields the Bloch equation 
                 in the presence of spin-orbit interaction: 
                  \begin{equation}
      \partial_t {\bf \hat{s}}  =  {\bf \hat{s}} \times {\bf B}_{\rm SO} ({\bf p}),
      \end{equation}
     so that  the spin performs a  precession around  the momentum dependent spin-orbit field 
  ${\bf B}_{\rm SO} ({\bf p}) $. It is important to note,  that the spin-orbit field does not break 
   the invariance under  time reversal  ( ${\bf \hat{s}} \rightarrow - {\bf \hat{s}} , {\bf p} \rightarrow - {\bf p}$ ),
    in contrast to an external magnetic field ${\bf B }$.
    Therefore, 
    averaging over all directions of momentum, 
    there is no  spin polarization of the conduction electrons. However, 
                 injecting a spin-polarized electron with given momentum ${\bf p}$
                  into  a translationally  invariant wire,  its spin precesses 
                   in the spin-orbit field as the electron moves through the wire. 
                   The spin will be oriented again in the initial direction after it moved
                      a length $L_{\rm SO}$, the spin precession length. The precise magnitude of 
                       $ L_{\rm SO}$ does not only 
                    depend on the strength of the spin-orbit interaction but may also depend
                     on the direction of its movement  in the crystal, as we will 
                    discuss  below. 
                           
     \subsubsection{Spin Diffusion Equation}
          
        Translational invariance is broken by 
        the presence of  disorder due to  impurities and  lattice imperfections in the conductor.
           As the electrons scatter from the disorder potential elastically, their 
            momentum changes  in a stochastic way, resulting in 
             diffusive motion.  That results in a 
              change of the  the local  electron density $\rho({\bf r}, t) = \sum_{\alpha = \pm}
              \mid \psi_{\alpha} ({\bf r}, t) \mid^2 $, where $\alpha = \pm$  denotes the 
               orientation of the electron spin,  and $\psi_{\alpha} ({\bf r}, t)$ is the 
                position and time dependent electron wave function amplitude.
              On length scales exceeding the 
             elastic mean free path $l_e$, 
                 that density is governed by the 
              diffusion equation 
              \begin{equation} \label{diff}
                \frac{\partial {\rho}}{\partial t} = D_e {\bf \nabla}^2 {\rho} ,
              \end{equation}
              where the diffusion constant $D_e$ is related to the elastic scattering 
               time $\tau$ by $D_e = v_{\rm F}^2 \tau/d_D$, where $v_{\rm F}$ is the 
                Fermi velocity, and $d_D$ the Diffusion  dimension of the electron system. 
                That diffusion constant is related to the mobility of the electrons, $\mu_e = e \tau/m^*$ by 
                 the Einstein relation $\mu_e \rho = e 2 \nu D_e$, where $\nu$ is the density of states per spin
                  at the Fermi energy $E_{\rm F}$.
                   Injecting an electron at position 
                     ${\bf r_0}$ into a conductor with previously constant electron 
                   density $\rho_0$, the solution of the diffusion equation yields
                   that the electron density spreads in space according to  $
                   \rho ({\bf r}, t) = \rho_0 + \exp (- ({\bf r - r_0})^2/4 D_e t)/(4 \pi D_e t)^{d_D/2}$, where
                    $d_D$ is the dimension of diffusion. That dimension 
                     is equal to the kinetic dimension $d$, $d_D=d$, if the elastic mean free path $l_e$ is smaller
                     than the size of the sample in all directions. If the elastic mean free path 
                     is larger than the sample size in one direction the diffusion dimension reduces by one, accordingly.
                      Thus, on average   the variance of the distance the  electron moves after time $t$ is 
                      $ \langle ({\bf r -r_0})^2 \rangle = 2 {d_D} D_e t$. This introduces a new length scale, 
                       the diffusion length $L_{D} (t) = \sqrt{ D_e t}$.
We can rewrite the density as $\rho=   \langle \psi^{\dagger}({\bf r},t ) \psi ({\bf r},t ) \rangle,$
                              where $\psi^{\dagger} = (\psi_+^{\dagger},\psi_-^{\dagger})$ is the two-component vector of the up (+), and down (-) spin 
 fermionic creation operators, and $\psi$ the 2-component vector of annihilation operators, 
 respectively, $\langle\ldots\rangle$ denotes the expectation value. 
  Accordingly, the spin density ${\bf s} ({\bf r}, t) $ is expected to  satisfy a diffusion 
            equation, as well. The spin density is defined by 
                    \begin{equation} \label{spindensity}
 \mathbf{s}({\bf r},t) = \frac{1}{2} \langle \psi^{\dagger}({\bf r},t ){\mathbf \sigma }  \psi ({\bf r},t ) \rangle,
 \end{equation}
  where ${\mathbf \sigma } $ is the vector of Pauli matrices, 
\begin{equation*}
\sigma_x = \left( \begin{array}{rr}
            0  & 1 \\
            1 & 0 
         \end{array} \right), \sigma_y = \left( \begin{array}{rr}
            0  & -i \\
            i & 0 
         \end{array} \right),\text{ and } \sigma_z = \left( \begin{array}{rr}
            1 & 0 \\
            0 & -1 
         \end{array} \right).
\end{equation*}
  Thus the z-component  of the spin density is
   half the difference between the density of  spin up and down electrons, 
    $s_z = (\rho_+ - \rho_-)/2$, which is the local spin polarization of the electron system.   
   Thus, we can directly infer the   diffusion equation for $s_z$, 
    and, similarly, for the other components of the spin density, yielding, 
     without magnetic field  and spin-orbit interaction,\cite{torrey}
          \begin{equation}
          \frac{\partial {\bf s} }{\partial t} = D_e {\bf \nabla}^2 {\bf s} 
        - \frac{\bf s}{ \hat{\tau}_s}.
   \end{equation}
       Here, in the spin relaxation term we introduced the tensor   $\hat{\tau}_s$,
       which can have non-diagonal matrix elements. In the case of a diagonal matrix,  $\tau_{s x x} = \tau_{s yy} = \tau_2$, is  the spin dephasing time, and $ \tau_{s zz} = \tau_1$  the  spin relaxation time.
   The spin diffusion equation   can be written as a continuity equation for the spin density vector, 
  by defining  the spin diffusion  current  of the spin components $s_i$, 
    \begin{equation}
{\bf J}_{s_i} =  - D_e {\bf \nabla}  s_i.
   \end{equation}
    Thus, we get the continuity equation for the spin density components $s_i$, 
   \begin{equation}
     \frac{\partial { s_i}}{\partial t} +  {\bf \nabla} {\bf J}_{s_i}  = - \sum_j  \frac{ s_j}{ \tau_{s ij}}.
   \end{equation}
   
      \subsubsection{Spin-Orbit Interaction in Semiconductors}
   
\begin{figure}[ht]
\begin{center}
\epsfig{file=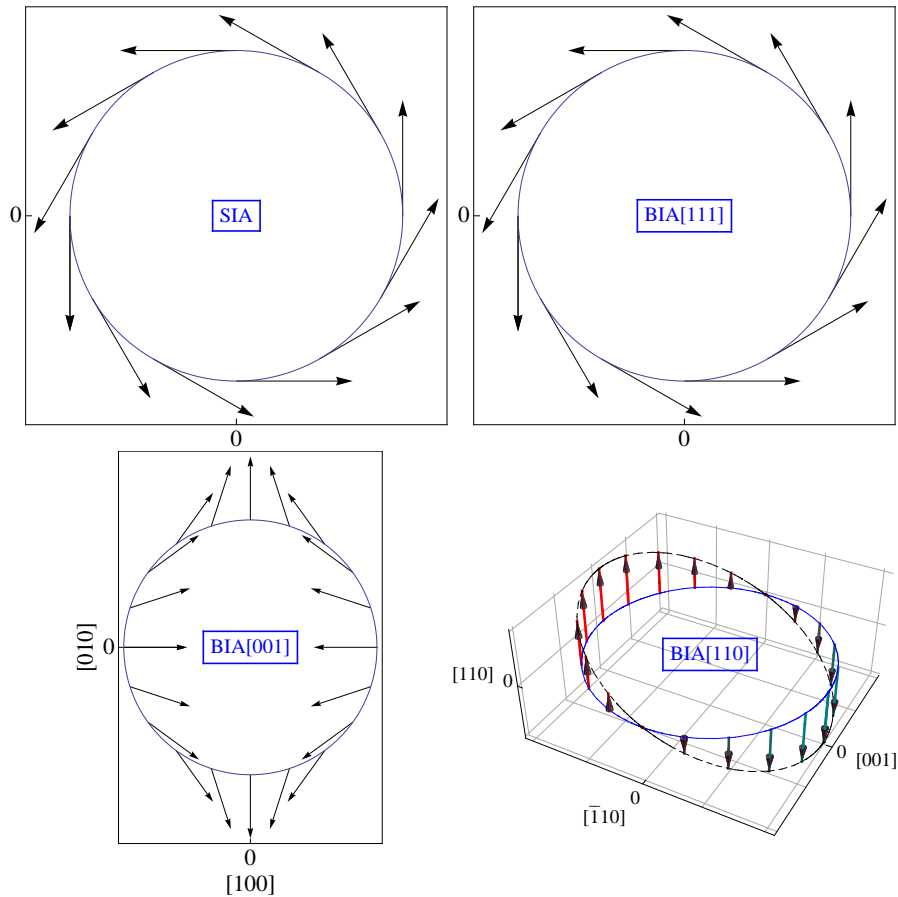,width= 12cm}
 \caption{ The spin-orbit vector fields for linear structure inversion asymmetry (Rashba) coupling, and for linear bulk inversion asymmetry (BIA) spin orbit coupling for quantum wells grown in [111], [001] and [110] direction, respectively.
} \label{Figspinorbitfield}
\end{center}
\end{figure}

    While silicon and germanium have in their diamond structure an inversion symmetry 
      around every 
          midpoint on each line connecting nearest neighbor atoms, this is not the case for
       III-V-semiconductors like  GaAs, InAs,  InSb, or ZnS.  These  have a zinc-blende 
        structure which can be obtained from a diamond structure with neighbored sites
         occupied by  the two different elements.  
        Therefore the  inversion symmetry is broken, which results in spin-orbit coupling. 
         Similarly, that symmetry is broken in II-VI-semiconductors. 
         This bulk inversion asymmetry (BIA)
      coupling, or often so called   Dresselhaus-coupling,   is anisotropic, as given by
     \cite{dresselhaus}
\begin{equation}
H_{\rm D} = \gamma_{D}   \left[ \sigma_x k_x (k_y^2 - k_z^2) + \sigma_y k_y (k_z^2 - k_x^2)+ \sigma_z k_z (k_x^2 - k_y^2)\right],
\label{Dresselhaus}
\end{equation}
where
$\gamma_{D}$ is the Dresselhaus-spin-orbit coefficient.
 Confinement  in quantum wells with  width $a$ on the order of the Fermi wave length 
  $\lambda_F$ yields accordingly a  spin-orbit interaction 
  where the momentum in growth direction is of the order of $1/a$.  Because of the 
   anisotropy of the Dresselhaus term, the spin-orbit interaction depends strongly 
    on the growth direction of the quantum well. 
     Grown in $[ 001 ]$ direction, one gets, taking the expectation value of 
      Eq.\,(\ref{Dresselhaus}) in the direction normal to the
     plane, noting that  $\langle k_z \rangle =  \langle k_z^3  \rangle =0$, 
      \cite{dresselhaus}
\begin{equation} \label{lineardressel}
H_{\rm D [001] } = \alpha_1 ( -\sigma_x k_x + \sigma_y k_y)
+ \gamma_D  ( \sigma_x k_x k_y^2 - \sigma_y k_y k_x^2).
\end{equation}
where  $\alpha_1 = \gamma_D \langle k_z^2 \rangle$ is the linear Dresselhaus parameter.
 Thus, inserting an electron with momentum along the 
  x-direction, with its spin initially polarized in z-direction, it will precess around the 
   x-axis as it moves along. 
  For narrow quantum wells, where 
    $\langle k_z^2 \rangle \sim 1/a^2 \ge k_F^2$  the linear  term  exceeds the cubic 
    Dresselhaus terms. 
    A special situation arises for quantum wells grown in the $[110]$- direction,
     where it turns out that the spin-orbit field is pointing normal to the 
      quantum well, as shown in Fig.\,\ref{Figspinorbitfield},
      so that an electron whose spin is initially polarized along the 
       normal of the plane, remains polarized as it moves in the quantum well. 
      In quantum wells with  asymmetric electrical confinement the inversion symmetry
       is broken as well. This structural inversion asymmetry (SIA) can be deliberately 
        modified by changing the confinement potential by application of a gate voltage.
         The resulting spin-orbit coupling,
       the SIA coupling, also  called Rashba-spin-orbit interaction\cite{rashba60} 
       is given by 
       \begin{equation} \label{rashba}
H_{\rm R} = \alpha_2  ( \sigma_x k_y - \sigma_y k_x),
\end{equation}
where $\alpha_2$ depends on the asymmetry of the confinement 
 potential $V(z)$ in the direction $z$, 
  the growth direction of the quantum well,  and can thus be deliberately changed by application of a gate potential.
 At first sight it looks as if the expectation value of the electrical field
  $\mathcal{E}_c = - \partial_z V(z)$ in the conduction band  state 
  vanishes, since the ground state of the quantum well must be  symmetric in $z$. 
  Taking into account  the coupling  to the  valence band,\cite{Lassnig1985,winklerbook}  
  the discontinuities in the effective mass,\cite{Malcher1986}
  and corrections due  to the coupling to  odd excited states,\cite{Bernardes2006}
   yields a sizable coupling parameter depending on the 
   asymmetry of the confinement potential\cite{winklerbook,fabianbook}.
 
       
    This dependence     allows one, in principle, 
        to control the electron spin with a gate potential, which can therefore be used as the basis of 
         a spin transistor.\cite{dattadas90}
       
\begin{figure}[ht]
\begin{center}
\epsfig{file=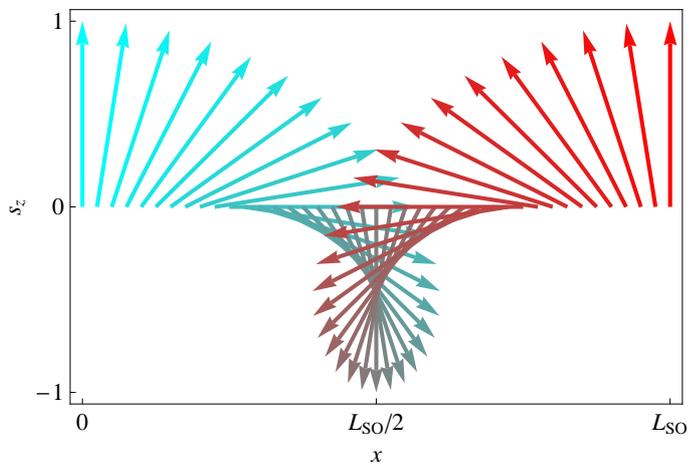,width= 9cm}
 \caption{ Precession of a spin injected at $x=0$, polarized in z-direction, as it moves by one
  spin precession length $L_{\rm SO}= \pi/m^* \alpha$ through the wire with linear Rashba spin orbit coupling $\alpha_2$.  }
 \label{Figballistic}
\end{center}
\end{figure}

       We can combine all spin-orbit couplings by introducing  the 
  spin-orbit field  such that the Hamiltonian has the form of a Zeeman term: 
  \begin{equation}
               H_{\rm SO} =  -  {\bf {s}} {\bf B}_{\rm SO} ({\bf k}) ,
               \end{equation} 
               where the spin vector is $ {\bf s}  = {\mathbf \sigma } /2$.
                But we stress again that since $ {\bf B}_{\rm SO} ({\bf k})  \rightarrow  {\bf  B}_{\rm SO} ({\bf - k})
                 = -  {\bf B}_{\rm SO} ({\bf k})  $  under the time reversal operation, 
                  spin-orbit coupling does not break time reversal symmetry, 
                   since the time reversal operation also changes the 
                    sign of the spin, $ {\bf s } \rightarrow - {\bf s}$. 
              Only an external magnetic field ${\bf B}$  breaks the time reversal 
               symmetry. 
                Thus, the electron spin operator ${\bf \hat{s} }$ 
                is for fixed electron momentum ${\bf k}$  governed by the 
                Bloch equations with  the spin-orbit field, 
                \begin{equation} \label{ballistic}
          \frac{\partial {\bf \hat{s}}}{\partial t} = {\bf \hat{s}} \times   \left( {\bf B + B_{\rm SO} ({\bf k}) } \right) 
- \frac{1} {\hat{\tau}_s} {\bf \hat{s}}.
   \end{equation}
    The spin relaxation tensor is no longer necessarily diagonal in the presence of spin-orbit interaction.\\
 In narrow  quantum wells where the cubic Dresselhaus coupling is weak compared to the linear Dresselhaus and  Rashba couplings,  the spin-orbit field is given by
             \begin{equation} \label{bso}
         {\bf B}_{\rm SO}({\bf k}) = -  2   \left( \begin{array}{c}
            -\alpha_1 k_x + \alpha_2 k_y \\
            \alpha_1 k_y - \alpha_2 k_x \\  0 
         \end{array} \right), 
         \end{equation}
        which changes both its direction and its amplitude   $\mid {\bf B}_{\rm SO}({\bf k}) \mid = 2 \sqrt{(\alpha_1^2 + \alpha_2^2)k^2 - 4 \alpha_1 \alpha_2 k_x k_y}$, 
         as the direction of the momentum ${\bf k}$ is changed.
          Accordingly, the 
          electron
           energy dispersion close to the Fermi  energy is in general anisotropic as given by 
            \begin{equation} \label{dispersionlinear}
            E_{\pm } = \frac{1}{2m^*} { k}^2 \pm \alpha k \sqrt{1-4 \frac{\alpha_1 \alpha_2}{\alpha^2} \cos \theta \sin \theta},
            \end{equation}
            where $k = \mid {\bf k} \mid$, $\alpha = \sqrt{\alpha_1^2 + \alpha_2^2}$, 
            and $k_x = k \cos \theta$. 
Thus, when an electron is injected with energy $E$, with momentum
              $k$ along the $[100]$-direction,  $k_x = k, k_y=0$, 
              its wave function is a superposition of plain waves with the positive 
               momenta $k_{\pm} = \mp \alpha m^* + m^* (\alpha^2 + 2E/m^*)^{1/2}$.
               The momentum difference $k_- - k_+ = 2 m^* \alpha$ causes 
               a rotation of the electron eigenstate in the spin subspace. 
                When at $x=0$ the electron spin was polarized up spin, 
                 with the Eigenvector
\begin{equation*}
 \psi(x=0) = \left( \begin{array}{r}
            1 \\
            0  
         \end{array} \right),
\end{equation*}
then, when its momentum points in x-direction, at a distance $x$, it will have rotated the spin as described by the Eigenvector 
\begin{equation}
         \psi(x) = \frac{1}{2}\left( \begin{array}{c}
            1\\
            \frac{\alpha_1 + i \alpha_2}{\alpha} 
         \end{array} \right) e^{i k_+ x} + \frac{1}{2}\left( \begin{array}{c}
            1\\
        -\frac{\alpha_1 + i \alpha_2}{\alpha}  
         \end{array} \right) e^{i k_- x}.
         \end{equation} 
          In Fig.\,\ref{Figballistic} we plot the corresponding spin density
          as defined in Eq.\,(\ref{spindensity}) for pure Rashba coupling,
          $\alpha_1 =0$.
         The spin will point again in the initial 
         direction, when the phase difference between the two plain waves
          is $2 \pi$, which gives the condition for  spin precession length as
           $2 \pi   = ( k_- - k_+) L_{\rm SO} $, 
           yielding for linear Rashba and Dresselhaus coupling, and the electron moving in 
           $[100]$- direction,
           \begin{equation} \label{lso}
           L_{\rm SO} = \pi/m^* \alpha.
           \end{equation}
            We note that the period of spin precession changes with the 
             direction of the electron momentum since the 
              spin-orbit field, Eq.\,(\ref{bso}), is anisotropic.
   \subsubsection{Spin Diffusion  in the Presence of Spin-Orbit Interaction}
 \label{spindiffusion}
\begin{figure}[ht]
\begin{center}
\epsfig{file=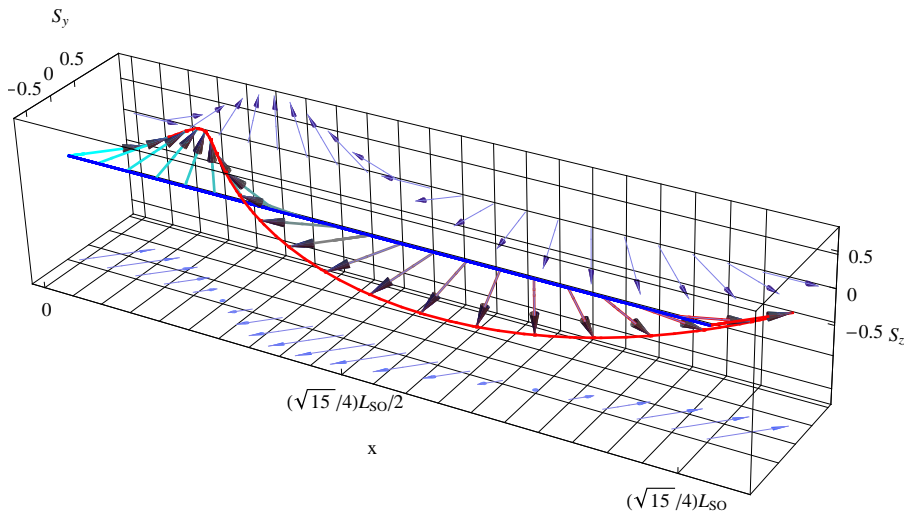,width= 12cm}
 \caption{ The spin density for linear Rashba coupling which is a solution of the 
  spin diffusion equation with the relaxation rate $ 7/16 \tau_s$. 
  The spin points  initially in the $x-y$-plane in the direction $(1,1,0)$. }
 \label{disspin}
\end{center}
\end{figure}

\begin{figure}[ht]
\begin{center}
\epsfig{file=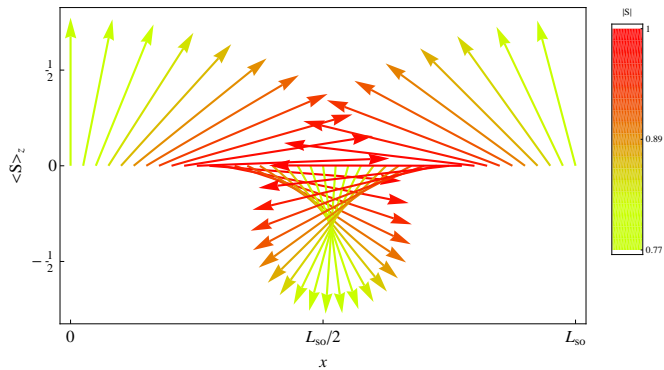,width= 9cm}
 \caption{ The spin density for linear Rashba coupling which is a solution of the 
  spin diffusion equation with the relaxation rate $1/\tau_s =  7/16 \tau_{s 0}$.
   Note that, compared to the ballistic spin density, Fig.\,\ref{Figballistic},  
    the period is slightly enhanced by a factor $4/\sqrt{15}$.  Also, the amplitude of the 
     spin density changes with the position $x$, in contrast to the ballistic case.  The color
      is changing in proportion to the spin density amplitude.
  }
 \label{disspinz}
\end{center}
\end{figure}

 As the electrons are scattered by imperfections like impurities and dislocations, their
  momentum is changed randomly. Accordingly, the direction of the spin-orbit field 
  $ {\bf B}_{\rm SO} ({\bf k}) $
   changes randomly as the electron moves through the sample. 
     This has two consequences:  the electron spin 
     direction becomes randomized, dephasing the spin precession and relaxing the spin polarization. 
     In addition, the spin precession term is modified, as the momentum ${\bf k}$
      changes randomly,   and has no longer the form 
     given in the ballistic Bloch-like equation, Eq.\,(\ref{ballistic}).
  One  can derive the diffusion equation for the expectation value of the 
   spin, the spin density Eq.\,(\ref{spindensity}) semiclassically, \cite{malshukov,raimondi}
    or by diagrammatic expansion.\cite{wenk2010}
 In order to get a better understanding on the meaning of this equation, we
  will give a simplified classical derivation, in the following. 
   The spin density at time $t  + \Delta t$ can be related to the one at the earlier
    time $t$. 
      Note that for ballistic times $\Delta t \le \tau $,
       the distance the electron has moved with a probability 
       $p_{\Delta {\bf x}}$, 
       $\Delta {\bf x}$, is  related to that time
        by the ballistic equation, $\Delta  {\bf x} =  {\bf k}(t) \Delta t/m$ when  the 
         electron moves with the momentum ${\bf k}(t)$.
         On this time scale   the spin evolution is still 
       governed by the  ballistic Bloch equation Eq.\,(\ref{ballistic}).
        Thus, we can relate the spin density at the position ${\bf x}$    at the time $t + \Delta t$,
         to the one at the earlier time $t$ at position ${\bf x} - \Delta {\bf x} $:
               \begin{equation} \label{expansion}
      {\bf s} ({\bf x}, t + \Delta t) = \sum_{\Delta {\bf x}}  p_{\Delta  {\bf x} }
      \left( \left(1 - \frac{1}{\hat{\tau}_s}   \Delta t\right)   {\bf s} ({\bf x} - \Delta {\bf x} , t )  - \Delta t  \left[ {\bf B}   + {\bf B}_{\rm SO} \left({\bf k}(t)  \right)    \right] \times {\bf s} ({\bf x}  - \Delta {\bf x} , t )  
       \right).
      \end{equation}
  Now, we can expand in $\Delta t$ to first order and in $\Delta {\bf x}$ to second order.
   Next, we average over the 
   disorder potential, 
    assuming that the electrons are scattered  isotropically, and 
     substitute $ \sum_{\Delta {\bf x}} p_{\Delta {\bf x} }\ldots= 
   \int (d \Omega/\Omega)\ldots$  where $\Omega$ is the total angle, and $ \int d \Omega$ denotes the
    integral over all angles with $\int (d \Omega/\Omega) = 1$.
     Also, 
     we get  $ \left( {\bf s} ({\bf x}, t + \Delta t) -   {\bf s} ({\bf x}, t)\right)/ \Delta t  \rightarrow \partial_t {\bf s} ({\bf x}, t)$ for $\Delta t \rightarrow 0$, 
      and $\langle  \Delta x_i^2  \rangle = 2 D_e \Delta t$, where $D_e$ is the diffusion constant. 
       While the disorder average yields $\langle \Delta {\bf x} \rangle = 0$, and 
        $ \langle {\bf B}_{\rm SO}( {\bf k} (t)) \rangle =0$, separately, for isotropic impurity scattering, 
      averaging their product
       yields a finite value, since   $\Delta {\bf x}$ depends on the momentum at time $t$,   $ {\bf k} (t)$, 
       yielding 
       $  \langle \Delta {\bf  x} { B}_{\rm SO i} \left({\bf k} (t) \right)    \rangle  =
          2 \Delta t \langle {\bf v}_{\rm F} {  B}_{\rm SO i} ({\bf k}(t))   \rangle $,
          where $\langle\ldots\rangle$ denotes the average over the Fermi surface. 
          This way, we can also evaluate the average of the spin-orbit term in Eq.\,(\ref{expansion}),
           expanded to first order in $ \Delta {\bf x} $, 
           and get, substituting $\Delta t \rightarrow \tau$ the spin diffusion equation, 
\begin{equation} \label{diffusive}
          \frac{\partial {\bf s}}{\partial t} =-   {\bf B} \times   {\bf s }  + D_e {\bf \nabla}^2 {\bf s} 
           + 2 \tau  \langle ( \mathbf{\nabla}  {\bf v_F}) {\bf B}_{\rm SO} ({\bf {p}}) \rangle 
          \times {\bf s} 
- \frac{1} {\hat{\tau}_s} {\bf s},
   \end{equation}
    where $\langle\ldots\rangle$ denotes the average over the Fermi surface. 
            Spin polarized electrons injected into the sample  spread  diffusively, 
            and  their spin polarization, while spreading diffusively as well, 
             decays  in amplitude exponentially in time. Since, between scattering 
              events the spins precess around the spin-orbit fields, one expects also 
               an oscillation of the polarization amplitude in space.  
             One can find the spatial distribution of the spin density which
              is the solution of Eq.\,(\ref{diffusive}) with the smallest decay
             rate      $\Gamma_s$. As an example, the solution  for linear Rashba coupling is, 
              \cite{raimondi}
              \begin{equation} \label{diffsol}
              {\bf s} ({\bf x}, t ) =  \left( \hat{e}_q  \cos {\bf q x } 
              + A  \hat{e}_z \sin {\bf q x}   \right) e^{- t/\tau_s},
              \end{equation}
              with $1/\tau_s  = 7/16 \tau_{s 0}$ where $1/\tau_{s 0} = 2 \tau k_F^2 \alpha_2^2$ and 
               where the amplitude of the momentum ${\bf q}$ is determined by 
              $D_e q^2 = 15/16 \tau_{s 0}$,
               and $A= 3/\sqrt{15}$, and $\hat{e}_q  = {\bf q}/q$.
                 This solution is plotted in Fig.\,\ref{disspin} for $\hat{e}_q =
                 (1,1,0)/\sqrt{2}$. In Fig.\,\ref{disspinz} we plot the linearly independent solution obtained
                  by interchanging $ \cos $ and $ \sin$  in Eq.\,(\ref{diffsol}), with the spin
                   pointing in z-direction, initially. We choose
                 $\hat{e}_q =
                 \hat{e}_x$. Comparison with the ballistic precession of the spin, Fig \ref{disspinz}
                  shows that the period of precession is enhanced by the factor $4/\sqrt{15}$
                   in the diffusive wire, and that the amplitude of the spin density is modulated,
                    changing from $1$ to $A= 3/\sqrt{15}$.

Injecting a spin-polarized electron at one point, say ${\bf x} =0$, 
its density spreads the same way it does without spin-orbit interaction, $\rho ({\bf r}, t) = \exp (- r^2/4 D_e t)/(4 \pi D_e t)^{d_D/2}$, where $r$ is the distance to the injection point. However, the decay of the spin density  is  periodically  modulated as a function of  $2 \pi \sqrt{15/16} r/L_{\rm SO} $.\cite{froltsov}
The spin-orbit interaction together with the   scattering from impurities is 
                      also a source of spin relaxation, as we discuss in the 
                       next Section together with other mechanisms of spin relaxation.
       We can find the classical spin diffusion  current in the presence of spin-orbit interaction,
       in a similar  way as  one can derive the classical diffusion 
        current: 
         The current at the position ${\bf r}$ is a sum 
          over all currents in its vicinity which are directed towards 
           that position. Thus, 
                   ${\bf j} ({\bf r},t) = \langle {\bf v} \rho ({\bf r} - \Delta {\bf x}) \rangle$
         where an angular average over all possible directions of the velocity ${\bf v}$ is taken.
          Expanding in $\Delta {\bf x} = l_e {\bf v}/v$, 
           and noting that  $\langle {\bf v}  \rho ({\bf r}) \rangle =0$,  one gets 
           ${\bf j} ({\bf r},t) = \langle {\bf v} (- \Delta {\bf x} ) \mathbf{\nabla} \rho ({\bf r}) \rangle
            = -(v_F l_e/2)  \mathbf{\nabla} \rho ({\bf r})  = - D_e \mathbf{\nabla} \rho ({\bf r}) $.
             For the classical spin diffusion current of  spin component $S_i$, as defined by  
             ${\bf j}_{S_i} ({\bf r},t) = {\bf v} S_i ({\bf r} ,t)$, there is the complication that 
              the spin keeps precessing as it moves from ${\bf r} - \Delta {\bf x}$ to 
             $  {\bf r} $, and that the spin-orbit field changes its direction with the direction of the 
              electron velocity ${\bf v}$. Therefore, the 0-th order term in the expansion 
              in $\Delta {\bf x} $ does not vanish, rather, we get 
               ${\bf j}_{S_i} ({\bf r},t) = \langle {\bf v} S_i^{\bf k} ({\bf r} ,t) \rangle - 
               D_e  \mathbf{\nabla} S_i ({\bf r},t) $, 
                where  $S_i^{\bf k} $ is the part of the spin density which evolved from 
                 the spin density at ${\bf r} - \Delta {\bf x}$ moving with velocity ${\bf v}$
                  and momentum ${\bf k}$.
                 Noting that the spin precession on ballistic scales $t \le \tau$ is 
                  governed by the Bloch equation, Eq.\,(\ref{ballistic}), we
                  find by integration of Eq.\,(\ref{ballistic}), that $S_i^{\bf k} =-  \tau \left( {\bf B}_{\rm SO} ({\bf k})  \times {\bf S} \right)_i $
                 so that we can rewrite  the
                   first term yielding the total spin diffusion current as
\begin{equation} \label{spincurrent}
        {\bf j}_{S_i} = - \tau \langle  {\bf v}_F \left({\bf B}_{\rm SO} ({\bf k})  \times {\bf S} \right)_i \rangle
 -D_e \mathbf{\nabla} S_i.
\end{equation}
  Thus, we can rewrite the spin  diffusion equation 
   in terms of this spin diffusion current and get the continuity equation
\begin{equation}
   \frac{\partial {\bf s}_i}{\partial t}  = -  D_e  \mathbf{\nabla}   {\bf j}_{S_i} +
    \tau    \langle   \mathbf{\nabla} {\bf v}_F \left({\bf B}_{\rm SO} ({\bf k})  \times {\bf S} \right)_i \rangle- \frac{1} {\hat{\tau}_{sij}} {\bf s}_j.
\end{equation}
  It is important to note that in contrast to the continuity equation for the 
   density,  there are two additional terms, due to the spin orbit interaction. 
    The last one is the spin relaxation tensor which will be considered in detail in the next section. 
     The other term arises due to the fact that Eq.\,(\ref{diffusive}) contains a factor
      $2$ in front of the spin-orbit precession term, while  the spin diffusion current 
       Eq.\,(\ref{spincurrent})
       does not contain that factor. This has important physical consequences,
       resulting in the suppression of the spin relaxation rate in quantum wires and quantum 
        dots as soon as their lateral extension is smaller than the spin precession length 
         $L_{\rm SO}$, as we will see in the subsequent Sections.
\section{Spin Relaxation Mechanisms}\label{sec:SpinRelaxMech}
          The intrinsic spin-orbit interaction itself causes the spin of the electrons to precess coherently, as the electrons move through a conductor, defining the spin precession length $L_{\rm SO}$, 
           Eq.\,(\ref{lso}).
           Since impurities and dislocations in the conductor randomize the electron momentum, 
            the impurity scattering is transferred into a randomization 
             of the electron  spin  by the 
             spin-orbit interaction, which  thereby results in  spin dephasing and spin relaxation. 
              This results in a new length scale, the spin relaxation length, $L_s$, which 
               is related to the spin relaxation rate $1/\tau_s$ by 
               \begin{equation}
               L_s = \sqrt{D_e \tau_s}.
               \end{equation}
\subsection{ D'yakonov-Perel' Spin Relaxation}
\begin{figure}[ht]
\begin{center}
\epsfig{file=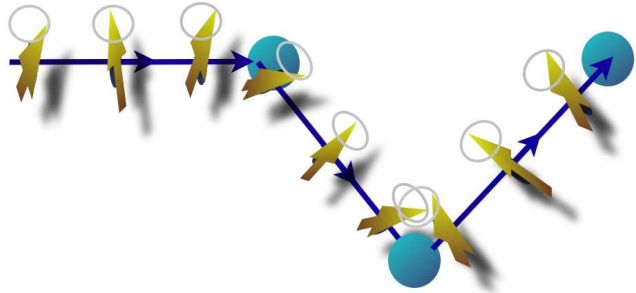,width= 9cm}
 \caption{ Elastic scattering from impurities changes the direction of the spin-orbit field
 around which the electron spin is precessing.
} \label{FigDP}
\end{center}
\end{figure}

\begin{figure}[ht]
\begin{center}
\epsfig{file=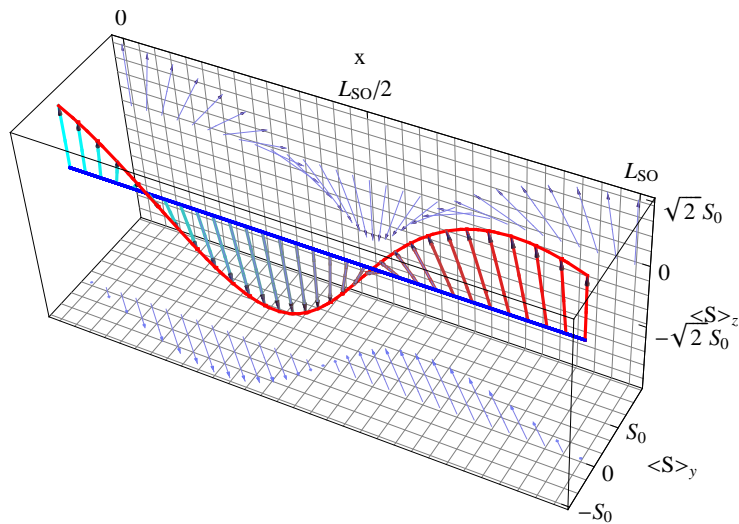,width= 12cm}
 \caption{ Persistent  spin helix solution of the spin diffusion equation for 
  equal magnitude of linear Rashba and linear Dresselhaus coupling, Eq.\,(\ref{Eq:spinhelix}).
} \label{Figspinhelix}
\end{center}
\end{figure}

          D'yakonov-Perel' spin relaxation (DPS) can be understood  qualitatively in the following way: 
             The spin-orbit field ${\bf B}_{\rm SO}({\bf k})$ changes its direction randomly 
              after each elastic scattering event from an impurity, that is, after a time 
               of the order of the elastic scattering time  $\tau$, 
                when the momentum is changed randomly as sketched in Fig.\,\ref{FigDP}.
Thus,  the spin has the time $\tau$ to
                 perform a precession around the present  
                  direction of the spin-orbit field, and can thus change its direction only by an angle 
                   of the order of $ {\bf B}_{\rm SO} \tau $ by precession.
                   After a  time $t$ with   $N_t = t/\tau$ scattering events, 
                    the direction of the spin will therefore 
                    have changed 
                 by an  angle of the order of  
                 ${\mid B_{\rm SO}\mid}  \tau \sqrt{N_t} = {\mid B_{\rm SO} \mid } \sqrt{\tau t}$.
                  Defining the spin relaxation time  $\tau_s$ as the time by which 
                   the spin direction has changed by an angle of order one, 
                    we thus find that $ 1/\tau_s \sim \tau  \langle  {\bf B}_{\rm SO} ({\bf k})^2
                    \rangle$, where the angular brackets denote integration over all angles.
                     Remarkably, this spin relaxation rate becomes smaller, the 
                      more scattering events take place, because the smaller the elastic scattering time 
                      $\tau$ is, the less time the spin has to change its direction by precession. 
                  Such a behavior is also  well known as {\it motional}, or {\it dynamic narrowing}
                   of  magnetic resonance lines.\cite{bloembergen}
     A more rigorous  derivation for the kinetic equation of the 
      spin density matrix  yields  
      additional interference terms, not taken into account in the above argument. 
        It can be obtained by iterating the expansion 
         of the spin density Eq.\,(\ref{expansion}) once in  the spin precession
       term, which yields
        the term 
        \begin{equation} \label{iterate}
        \left\langle  {\bf s}({\bf x},t) \times \int_0^{\Delta t} d t' {\bf B}_{\rm SO}({\bf k}(t'))
        \times \int_0^{\Delta t} d t'' {\bf B}_{\rm SO}({\bf k}(t''))
          \right\rangle,
        \end{equation}
         where $\langle\ldots\rangle$ denotes the average over all angles due to the scattering 
         from impurities. Since the electrons move ballistically at times smaller than 
          the elastic scattering time, the momenta are correlated only on time scales smaller
           than $\tau$, yielding $\langle k_i(t') k_j(t'') \rangle = (1/2) k^2 \delta_{i j} \tau 
           \delta (t' - t'')   $.\\ 
Noting that $({\bf A} \times {\bf B} \times {\bf C})_m = 
           \epsilon_{ijk} \epsilon_{klm} A_i B_j C_l$ and $\sum    \epsilon_{ijk} \epsilon_{klm}
            = \delta_{il} \delta_{jm} - \delta_{im}\delta_{jl}$ we find that 
             Eq.\,(\ref{iterate}) simplifies to $- \sum_i(1/\tau_{si j}) S_j$,
              where the matrix elements of the spin relaxation terms are given by,
         \cite{dyakonov72_1}
        \begin{equation} \label{dp}
        \frac{1}{\tau_{s  i j}} =  \tau    \left( \langle  {\bf B}_{\rm SO}({\bf k})^2  \rangle 
        \delta_{i j } -    \langle { B}_{\rm SO}({\bf k})_i    { B}_{\rm SO}({\bf k})_j  \rangle  \right),
        \end{equation}
 where $\langle\ldots\rangle$ denotes the average over the direction of the momentum 
  ${\bf k}$.     
   These non-diagonal terms can diminish the spin relaxation and even 
    result in vanishing spin relaxation. 
    As an example, we consider a quantum well where the linear Dresselhaus coupling for quantum wells  grown in $[ 001]$  direction, Eq.\,(\ref{lineardressel}),
      and linear Rashba-coupling, Eq.\,(\ref{rashba}), are the dominant spin-orbit couplings.
       The energy dispersion is anisotropic, as given by Eq.\,(\ref{dispersionlinear}),
        and the  spin-orbit field ${\bf B}_{\rm SO}({\bf k})$ changes its direction and
        its amplitude   with the direction of the momentum ${\bf k}$:
         \begin{equation} \label{bsok}
         {\bf B}_{\rm SO}({\bf k}) = -  2   \left( \begin{array}{c}
            -\alpha_1 k_x + \alpha_2 k_y \\
            \alpha_1 k_y - \alpha_2 k_x \\  0 
         \end{array} \right), 
         \end{equation}
          with $\mid {\bf B}_{\rm SO}({\bf k}) \mid = 2 \sqrt{(\alpha_1^2 + \alpha_2^2)k^2 - 4 \alpha_1 \alpha_2 k_x k_y}$.
                     Thus we find the spin relaxation tensor as, 
             \begin{equation} \label{spinrelaxation}
         \frac{1}{\hat{\tau}_s} (k)= 4  \tau k^2  \left( \begin{array}{ccc}
           \frac{1}{2} \alpha^2 & - \alpha_1 \alpha_2  & 0  \\
         -   \alpha_1 \alpha_2 &  \frac{1}{2} \alpha^2 & 0 \\ 
               0 &  0 &  \alpha^2
         \end{array} \right).
         \end{equation}
        Diagonalizing this matrix, one finds the three eigenvalues 
        $(1/\tau_s) (\alpha_1 \pm \alpha_2)^2/\alpha^2$ and 
        $2/\tau_s$ where $\alpha^2 = \alpha_1^2 + \alpha_2^2$, and 
           $1/\tau_s = 2 k^2 \tau \alpha^2$.
         Note, that one of these eigenvalues of the 
          spin relaxation tensor vanishes when $\alpha_1 = \alpha_2=\alpha_0$. 
            In fact, this is a special case, when the spin-orbit field does not change its direction 
             with the momentum: 
             \begin{align} 
             {\bf B}_{\rm SO}({\bf k}) \mid_{\alpha_1 = \alpha_2=\alpha_0} =&  2 \alpha_0
             (k_x - k_y)  \left( \begin{array}{r}
           1 \\
            1  \\ 
               0 
         \end{array} \right).\label{bso110}\\
            \intertext{In this case the constant spin density given by} 
             {\bf S} =& S_0  \left( \begin{array}{r}
           1 \\
            1  \\ 
               0 
         \end{array} \right),
        \end{align}
         does not decay in time, since the spin density vector is parallel  to the 
          spin orbit field ${\bf B}_{\rm SO}({\bf k})$, Eq.\,(\ref{bso110}), and cannot precess, as has been noted in Ref.\, [\onlinecite{averkiev}].
            It turns out, however,  that there are two more modes which do not decay in time, 
             whose spin relaxation rate vanishes for $\alpha_1 = \alpha_2$.
               These modes are not homogeneous in space, and correspond to 
                precessing spin densities. They were found previously in a numerical Monte Carlo
                simulation and 
   found not to decay in time, being called therefore {\it persistent spin helix}.\cite{bernevig,ohno}
     Recently, a  long living inhomogeneous spin density distribution has been 
      detected experimentally in Ref.\,[\onlinecite{spinhelix}].
   We can now  get   these {\it persistent spin helix modes}
    analytically, by  solving the 
                 full spin diffusion equation Eq.\,(\ref{diffusive}) with the spin relaxation tensor given by
                  Eq.\,(\ref{spinrelaxation}).
                   We can diagonalize that equation, noting that its eigenfunctions 
                    are plain waves ${\bf S} ({\bf x}) \sim \exp ( i {\bf Q  x } - E t)$. 
                  Thereby one finds, first of all,
                  the mode  with Eigenvalue  $E_1 =  D_e {\bf Q}^2$, 
                    with the spin density 
                        \begin{equation} \label{homogenous}
             {\bf S} = S_0  \left( \begin{array}{r}
           1 \\
            1  \\ 
               0 
         \end{array} \right)  \exp ( i {\bf Q  x } - D_e {\bf Q}^2 t).
        \end{equation}
          Indeed  for ${\bf Q} =0$, the homogeneous solution, it does not decay in time, in agreement with 
           the solution we found above, Eq.\,(\ref{homogenous}).
        There are, however,  two more modes with the eigenvalues
          \begin{equation}
          E_{\pm} =   \frac{1}{\tau_s} ({\bf \tilde{Q}}^2 +2 \pm 2 \mid   \tilde{Q}_x -  \tilde{Q}_y \mid ),
          \end{equation}
           where $ \tilde{Q} = L_{\rm SO} Q/2 \pi$.
           At  $\tilde{Q}_x = -\tilde{Q}_y = \pm 1$,  these modes do not decay in time. 
           These two stationary solutions, are   
              \begin{equation}\label{Eq:spinhelix}
           {\bf S} = S_0  \left( \begin{array}{c}
           1\\
           -1\\ 
            0 
         \end{array} \right)  \sin \left(   \frac{2 \pi}{L_{\rm SO}} (x-y ) \right)
         + S_0 \sqrt{2}  \left( \begin{array}{c}
           0 \\
            0  \\ 
               1 
         \end{array} \right)  \cos \left(   \frac{2 \pi}{L_{\rm SO}} (x-y ) \right),
           \end{equation}
            and the linearly independent solution, obtained by interchanging 
            $\cos$ and $\sin$ in Eq.\,(\ref{Eq:spinhelix}).
         The spin precesses as the electrons diffuse along the quantum wire 
            with the period $L_{\rm SO}$, the spin precession length, forming a 
             {\it persistent spin helix}, as shown in Fig.\,\ref{Figspinhelix}.
\subsection{ DP Spin Relaxation with Electron-Electron and Electron-Phonon Scattering}
             It has been noted, that the momentum scattering which limits the 
              D'yakonov-Perel' mechanism of spin relaxation 
              is not restricted to impurity scattering, but can also be due to electron-phonon 
               or electron-electron interactions.\cite{glazov,glazov2,punnoose,dysonridley} Thus the scattering time,
               $\tau$ is the total scattering time as defined by, 
               \cite{glazov,glazov2} $1/\tau = 1/\tau_0 + 1/\tau_{ee} + 1/\tau_{ep}$,
                where 
                $1/\tau_0 $ is the elastic scattering rate due to  scattering 
                 from impurities with potential $V$, given  by 
                  $1/\tau_0 = 2 \pi \nu n_i \int (d \theta/2 \pi)
                   (1- \cos \theta) \mid V({\bf k}, {\bf k'}) \mid^2$,
                   where
                    $\nu$ is the density of states per spin at the Fermi energy, 
                    $n_i$ is the concentration of impurities with potential 
                    $V$, and ${\bf k} {\bf k'} = k k' \cos (\theta)$. 
                     In degenerate semiconductors and in metals,
                      the electron-electron scattering rate is given by   the Fermi liquid  inelastic 
                       electron scattering rate $1/\tau_{ee} \sim T^2/\epsilon_F$. 
                        The electron-phonon scattering time $1/\tau_{ep} \sim T^5$ decays faster with 
                         temperature. Thus, at low temperatures the 
                          DP spin relaxation is dominated by elastic impurity scattering $\tau_0$. 
           \subsection{Elliott-Yafet Spin Relaxation}
         Because of the spin-orbit interaction the conduction electron wave functions
         are not Eigenstates of the electron spin, but have an admixture of both 
          spin up and spin down wave functions. Thus, a nonmagnetic impurity 
           potential $V$ can change the electron spin, by changing their momentum 
             due to the spin-orbit coupling. 
             This results in another source of spin relaxation which is stronger, the more often the electrons 
              are scattered, and is thus proportional to the momentum scattering rate
              $1/\tau$.\cite{elliott,yafet}
              For degenerate III-V semiconductors one finds\cite{chazalviel,pikustitkov}
              \begin{equation}
              \frac{1}{\tau_s} \sim \frac{\Delta_{\rm SO}^2}{(E_G + \Delta_{\rm SO})^2} \frac{E_{\bf k}^2}{E_G^2} \frac{1}{\tau ({\bf k})},
              \end{equation}
              where $E_G$ is the gap between the valence and the conduction band of the 
              semiconductor, $E_{\bf k}$ the energy of the conduction electron, and $\Delta_{\rm SO}$ is the spin-orbit splitting of the valence band.
               Thus, the Elliott-Yafet spin relaxation (EYS) can be   distinguished,
                being proportional to $1/\tau$, and thereby to the resistivity,
                in contrast to   the  DP spin scattering rate, 
                Eq.\,(\ref{dp}), which is proportional to the conductivity.
                 Since the EYS decays in proportion to the inverse of the band gap, 
                 it is negligible in  large band gap semiconductors like $Si$ and $GaAs$. 
                  The scattering rate $1/\tau$ 
                   is again the sum of the  impurity scattering rate,\cite{elliott}
                   the electron-phonon scattering rate,\cite{yafet,grimaldi} and electron-electron interaction,
                  \cite{boguslawski} so that all these scattering processes result in EYS.
                   In non-degenerate semiconductors, where the Fermi energy is below the 
                    conduction band edge, $1/\tau_s \sim \tau T^3/E_G$ attains a stronger temperature dependence. 
                    \subsection{ Spin Relaxation due to Spin-Orbit Interaction with  Impurities}
                The spin-orbit interaction, as defined in  Eq.  (\ref{so}), arises whenever there
                 is a gradient in an electrostatic potential. Thus, the impurity potential 
                 gives rise to the spin-orbit interaction 
                 \begin{equation}
                 V_{\rm SO} =  \frac{1}{2 m^2 c^2} \nabla V \times {\bf k} ~{\bf s}.
                 \end{equation}
                  Perturbation theory yields then directly the 
                   corresponding spin relaxation rate
                   \begin{equation}
                   \frac{1}{\tau_s} = \pi \nu n_i  \sum_{\alpha,\beta} \int \frac{d \theta}{2 \pi}
                   (1- \cos \theta) \mid V_{\rm SO}({\bf k}, {\bf k'})_{\alpha \beta} \mid^2,
                   \end{equation}
                    proportional to the concentration of impurities $n_i$. Here $\alpha, \beta = \pm$
                     denotes the spin indices. 
                    Since the spin-orbit interaction increases with the atomic number $Z$ of the 
                      impurity element, this spin relaxation increases as $
         Z^2$, being stronger for heavier element impurities.
\subsection{Bir-Aronov-Pikus Spin Relaxation}  
The exchange interaction $J$ between electrons and holes in  p-doped semiconductors results in spin relaxation,  as well.\cite{bir1,bir3} Its strength is proportional to the density of holes $p$ and depends  on their itinerancy.
             If the holes are localized they act like magnetic impurities. If they are itinerant, 
              the spin of the conduction electrons is transferred  by the 
               exchange interaction to the holes, where the spin-orbit splitting of the valence
                bands results in  fast spin relaxation of the hole spin due to the 
                 Elliott-Yafet, or the D'yakonov-Perel' mechanism.
           \subsection{Magnetic Impurities}
Magnetic impurities  have a spin ${\bf S}$ which interacts with the spin of the conduction electrons by the exchange interaction $J$, resulting in a spatially and temporarily fluctuating local magnetic field
              \begin{equation}
               {\bf B}_{\text{MI}} ({\bf r}) = -  \sum_i  J \delta ({\bf r - R}_i)  {\bf S},
              \end{equation}
where the sum is over the position of the magnetic impurities ${\bf R}_i$. This  gives rise to spin relaxation of the conduction electrons, with a rate  given by
\begin{equation} \label{tausm}
  \frac{1}{\tau_{\text{Ms}}} = 2 \pi n_M \nu J^2 S (S+1),
\end{equation}
where $n_M$ is the density of magnetic impurities, and $\nu$ is the density of states at the Fermi energy. Here, $S$ is the spin quantum number of the magnetic impurity, which can take the values $S= 1/2,1,3/2,2\ldots$. Antiferromagnetic  exchange interaction between the magnetic impurity spin and the conduction electrons  results in a competition between the conduction electrons to form a singlet with the impurity spin, which results in enhanced nonmagnetic and magnetic scattering. At low temperatures the magnetic impurity  spin is screened by the conduction electrons resulting in a  vanishing of the magnetic scattering rate. Thus, the spin scattering from magnetic impurities  has a maximum at a temperature of the order of the Kondo temperature $T_K \sim E_F \exp (-1/\nu J)$, where $\nu$ is the density of states at the Fermi energy.\cite{kondo,kondo1,kondo2} In semiconductors $T_K$ is exponentially small due to the small effective mass and the resulting small density of states  $\nu$. Therefore, the magnetic moments remain free at the experimentally  achievable temperatures.
                   At large concentration of magnetic impurities, the RKKY-exchange interaction between the magnetic impurities quenches however  the spin quantum dynamics, so that $S(S+1)$ is replaced by its classical value $S^2$. In Mn-p-doped GaAs, the exchange interaction between the Mn dopants and the holes can result in compensation of the hole spins  and therefore a suppression of the Bir-Aronov-Pikus (BAP) spin relaxation.\cite{molenkamp}
               \subsection{Nuclear Spins}
Nuclear spins interact by the hyperfine interaction
             with conduction electrons. The hyperfine interaction between nuclear spins ${\bf \hat{ I}}$ and the 
               conduction electron spin, ${\hat{s}}$, results in a local Zeeman field
               given by \cite{overhauser}
               \begin{equation} \label{hyperfine}
               {\bf \hat{B}_N({\bf r})} =-\frac{8 \pi}{3} \frac{g_0 \mu_B}{\gamma_g} \sum_n \gamma_n {\bf \hat{ I}},
               \delta({\bf r -R}_n),
               \end{equation}
                where $\gamma_n$ is the gyromagnetic ratio of the nuclear spin.
                 The spatial and temporal  fluctuations of this hyperfine interaction field 
                  result in spin relaxation 
                   proportional to its variance,  
                  similar to the spin relaxation by magnetic impurities.
              \subsection{Magnetic Field Dependence of Spin Relaxation}
            The  magnetic field changes the electron momentum due to the 
             Lorentz force, resulting in a  continuous change of the 
              spin-orbit field, which similar to the momentum scattering 
               results in motional narrowing and thereby a  reduction of DPS: \cite{ivchenko1,ivchenko2,pikustitkov,burkovbalents}
            \begin{equation} \label{DPSB}
            \frac{1}{\tau_s} \sim \frac{\tau}{1 + \omega_c^2 \tau^2}. 
            \end{equation}
             Another source of a magnetic field dependence is the precession around the 
              external magnetic field. In bulk semiconductors and  for magnetic fields
              perpendicular to a quantum well, the orbital mechanism is dominating, however.   
             This magnetic field dependence can be used to identify the spin relaxation mechanism, 
              since the EYS does have only  a weak magnetic field dependence due to the weak Pauli-paramagnetism.
\subsection{ Dimensional Reduction of Spin Relaxation}
       Electrostatic     confinement of conduction electrons
         can   reduce the effective dimension 
             of  their motion. 
             In {\it quantum dots}, the electrons are confined in all 
              three directions, and  the energy spectrum consists of 
               discrete levels like in atoms. Therefore, 
                the energy conservation  restricts relaxation processes severely, 
                 resulting in strongly enhanced spin relaxation times in quantum dots.\cite{khaetskii,af01}
                Then,  spin relaxation can only occur due to absorption or 
                 emission of phonons, yielding spin relaxation rates 
                  proportional to the inelastic  electron-phonon 
                 scattering rate.\cite{khaetskii} Quantitative comparison of the various spin relaxation
                  mechanisms  in GaAs quantum dots resulted in the conclusion that 
                   the spin relaxation is dominated by the hyperfine interaction.\cite{erlingsson,khaetskiinuclearprb,khaetskiinuclearprl}
           A similar conclusion can be drawn from    experiments on low temperature 
            spin relaxation in low density n-type GaAs, where the localization of the 
            electrons in the {\it impurity band} results in spin relaxation dominated by 
            hyperfine interaction as well.\cite{dzhioev,jalabert}
             For linear Rashba and linear Dresselhaus spin-orbit coupling
              we can see from the spin diffusion equation Eq.\,(\ref{diffusive})
               with the DP spin relaxation tensor Eq.\,(\ref{spinrelaxation})
                that  the spin relaxation vanishes, when the spin current 
                 Eq.\,(\ref{spincurrent}) vanishes, in which case the last two terms of
                  Eq.  (\ref{diffusive}) cancel exactly. 
     The vanishing of the spin current is imposed by 
                  hard wall boundary condition for which the 
                   spin diffusion current vanishes at the boundaries of the sample,
                    ${\bf j}_{S_i} {\bf n} \mid_{\rm Boundary} = 0$, where ${\bf n}$ is the normal 
                     to the boundary. When the quantum dot is smaller than the 
                      spin precession length $L_{\rm SO}$ the lowest energy mode thus corresponds to 
                       a homogeneous solution with vanishing spin relaxation rate. 
                         Cubic  spin-orbit coupling does not yield  such a vanishing of the DP spin 
                         relaxation rate. Only in quantum dots whose size does not exceed the 
                          elastic mean free path $l_e$  the DP spin relaxation from cubic
                           spin relaxation becomes diminished. 
In {\it quantum wires}, the electrons have a continuous spectrum of delocalized states.
         Still, transverse confinement can reduce the DP 
          spin relaxation as we review in the next section. 
\section{Spin-Dynamics in Quantum Wires}\label{sec:SpinDynQW}
               \subsection{ One-Dimensional  Wires} 
               \label{1D}
               In one dimensional wires, whose width $W$ is of the 
                order of the Fermi wave length $\lambda_F$, impurities can only reverse
                the momentum $p \rightarrow -p$. Therefore, 
                 the spin-orbit field can only change its sign, 
                  when a scattering from impurities occurs. 
                  ${\bf B}_{\rm SO}(p) \rightarrow {\bf B}_{\rm SO}(-p) = - {\bf B}_{\rm SO}(p)$.
                   Therefore, the precession axis and the amplitude of the 
                    spin orbit field  does not change, reversing only the 
                     spin precession, so that the 
                      D'yakonov-Perel'-spin relaxation
                       is absent in one dimensional wires.\cite{meyer}
                      In an external  magnetic field, the  precession around the 
                       magnetic field axis, due to the Zeeman-interaction is competing with 
                        the spin-orbit field, however. Then, as the electrons are scattered
                         from impurities, 
                          both the precession axis and the amplitude of the total precession field
                           is changing, since
\begin{equation*}
 \mid {\bf B} + {\bf B}_{\rm SO}(-p) \mid = \mid {\bf B} - {\bf B}_{\rm SO}(p)\mid \neq \mid {\bf B} + {\bf B}_{\rm SO}(p) \mid,
\end{equation*}
resulting in spin dephasing and relaxation, as the sign of the momentum changes randomly.
\subsection{Spin-Diffusion  in Quantum Wires}
\begin{figure}[ht]
\begin{center}
\epsfig{file=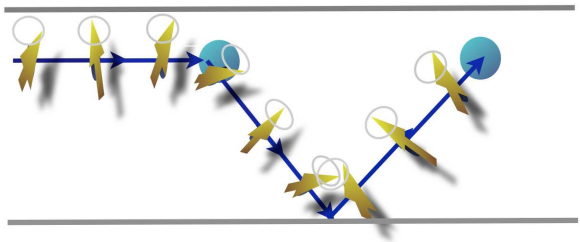,width= 8cm}
 \caption{ Elastic scatterings from impurities and from the 
  boundary of the wire change the direction of the spin-orbit field
 around which the electron spin is precessing.
} \label{FigDPw}
\end{center}
\end{figure}

\begin{figure}[ht]
\begin{center}
\epsfig{file=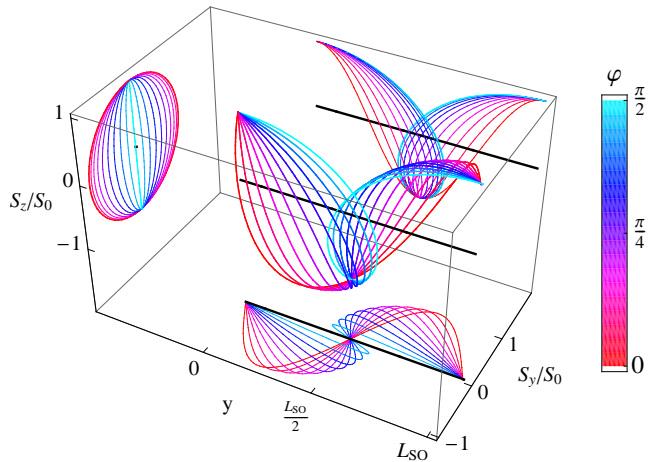,width= 12cm}
 \caption{ Persistent  spin helix solution of the spin diffusion equation 
 in a quantum wire whose width $W$ is smaller than the spin precession length 
 $L_{\rm SO}$
 for varying ratio 
 of linear Rashba $\alpha_2 = \alpha \sin \varphi $ and linear Dresselhaus coupling, 
 $\alpha_1 = \alpha \cos \varphi $,
 Eq.\,(\ref{Eq:spinhelix1d}), for  fixed $\alpha$ and $L_{\rm SO} = \pi/m^* \alpha$.
} \label{spinhelix1d}
\end{center}
\end{figure}

          How does the spin relaxation rate depend on the 
            wire width $W$ when the quantum wire 
         has  more than one channel occupied, $W > \lambda_F$?
           Clearly,  for large wire widths, the spin relaxation rate should converge
            to  a finite value, while it  vanishes for 
             $W \rightarrow \lambda_F$. It is both of practical importance for 
             spintronic applications and of fundamental interest to know on which length scales
              this crossover occurs. Basically, there are three intrinsic length scales 
               characterizing the quantum wire relative to its width $W$. 
                The Fermi wave length $\lambda_F$, the elastic mean free path $l_e$ and 
                 the spin precession length $L_{\rm SO}$, Eq.\,(\ref{lso}).
                Suppression of spin relaxation for wire widths not exceeding the elastic mean 
                free path $l_e$, 
               has been predicted and obtained numerically in Refs.\, 
               [\onlinecite{bournel,malshukov,kizelev,pareek,kaneko,dragom}]. Is  the  spin relaxation rate also suppressed in 
          {\it diffusive wires}   in which the elastic mean free path is smaller than the wire width
           as in the wire shown schematically in Fig.\,\ref{FigDPw}?
           We will answer this question by means of an analytical derivation in the following.
The transversal confinement imposes that the spin current vanishes
      normal to the boundary,   ${\bf j}_{S_i} {\bf n} \mid_{\rm Boundary} = 0$. 
       For a wire grown along the $[010]$ direction, ${\bf n} =  \hat{e}_x$ is the  unit
        vector in the  $x$-direction. 
        For wire widths $W$  smaller than the 
         spin precession length $L_{\rm SO}$, the 
          solutions with the lowest energy have thus a vanishing 
           transverse spin current, and  the spin diffusion equation Eq.\,(\ref{diffusive}) becomes
\begin{equation} \label{diffusivewire}
   \frac{\partial { s_i}}{\partial t}  = -  D_e  \partial_y   j_{S_i y} +
    \tau    \langle   \mathbf{\nabla} {\bf v}_F \left({\bf B}_{\rm SO} ({\bf k})  \times {\bf S} \right)_i \rangle - \sum_j \frac{1} {\hat{\tau}_{s ij}} { s_j}.
\end{equation}
 with                 
  \begin{equation} \label{spincurrentBC}
        j_{S_i  x}\mid_{x = \pm W/2} = \left( - \tau \langle   v_x \left({\bf B}_{\rm SO} ({\bf k})  \times {\bf S} \right)_i \rangle
 -D_e \partial_x S_i \right) \mid_{x = \pm W/2} =0,  
  \end{equation}
  where $W$ is the width of the wire. 
      One sees that this equation has a persistent solution,
       which does not decay in time and  is homogeneous 
       along the wire, 
       $\partial_y S =0$. In this special case the spin diffusion equation simplifies to\cite{raimondi} 
       \begin{equation}
            \partial_t {\bf S} = -
         \frac{1}{{\tau}_s \alpha^2}   \left( \begin{array}{ccc}
            \alpha_1^2 & - \alpha_1 \alpha_2  & 0  \\
         -   \alpha_1 \alpha_2 &  \alpha_2^2 & 0 \\ 
               0 &  0 &  \alpha^2
         \end{array} \right) {\bf S}.
         \end{equation}
          Indeed this has one persistent solution 
           given by
           \begin{equation}
             {\bf S} = S_0  \left( \begin{array}{c}
           \alpha_2 \\
            \alpha_1  \\ 
               0 
         \end{array} \right),
        \end{equation}
         Thus, we can conclude that the boundary conditions impose an effective alignment of all spin-orbit fields, 
          in a direction identical to the one it would attain in a one-dimensional wire, 
           along the [010]-direction, setting 
           $k_x=0$ in Eq.\,(\ref{bsok}),
              \begin{equation} \label{bsok1d}
         {\bf B}_{\rm SO}({\bf k}) = -  2 k_y  \left( \begin{array}{r}
             \alpha_2  \\
            \alpha_1  \\  0 ~
         \end{array} \right), 
         \end{equation}
          which therefore does not change its direction when the electrons are scattered. 
            This is remarkable, since this alignment already
            occurs in wires with many channels, where the impurity scattering is two-dimensional, and the transverse momentum $k_x$  actually can be finite.
              Rather,   the alignment 
               of the spin-orbit field, accompanied by a 
                suppression of the DP spin relaxation rate occurs due to the
                 constraint on the spin-dynamics imposed by the  boundary 
                  conditions  as soon as the wire width $W$ is smaller
               than the length  scale which governs the spin 
               dynamics, namely, the spin precession length $L_{\rm SO}$.
        It turns out that the spin diffusion equation Eq.\,(\ref{diffusivewire})
         has also two long persisting spin helix solutions in narrow wires\cite{kettemann,wenk2010} which oscillate periodically with the  period  $ L_{\rm SO} = \pi/m^* \alpha$.
           In contrast to the situation in 2D systems we reviewed in the previous Section, 
             in quantum wires of width $W < L_{\rm SO}$ these solutions  
                 are long persisting even for $\alpha_1 \neq \alpha_2$. 
                   These two stationary solutions, are   
              \begin{equation}\label{Eq:spinhelix1d}
           {\bf S} = S_0  \left( \begin{array}{c}
           \frac{\alpha_1}{\alpha} \\
            -\frac{\alpha_2}{\alpha}  \\ 
               0 
         \end{array} \right)  \sin \left(   \frac{2 \pi}{L_{\rm SO}} y \right)
         + S_0   \left( \begin{array}{c}
           0 \\
            0  \\ 
               1 
         \end{array} \right)  \cos \left(   \frac{2 \pi}{L_{\rm SO}} y \right),
           \end{equation}
            and the linearly independent solution, obtained by interchanging 
            $\cos$ and $\sin$ in Eq.\,(\ref{Eq:spinhelix1d}).
         The spin precesses as the electrons diffuse along the quantum wire 
            with the period $L_{\rm SO}$, the spin precession length, forming a 
             {\it persistent spin helix}, whose  x-component is proportional to the 
              linear Dresselhaus-coupling $\alpha_1$ 
               while its y-component is proportional to the Rashba-coupling
                $\alpha_2$ as seen in Fig.\,\ref{spinhelix1d}.
A similar reduction of the spin relaxation rate  is not effective for cubic spin-orbit coupling
                   for wire widths exceeding the elastic mean free path $l_e$.
                   One can derive the spin relaxation rate as function of the wire
                    width for diffusive wires  $l_e < W < L_{\rm SO}$.
        The total spin relaxation rate, 
         in the presence of both linear Rashba spin-orbit coupling $\alpha_2$ 
          and linear and cubic Dresselhaus coupling $\alpha_1$, and $\gamma_D$, 
           is as function of wire width  $W$ given by,\cite{kettemann}  
    \begin{equation}\label{SRW}
    \frac{1}{ \tau_s} (W) =  \frac{1}{12}\left(\frac{W}{L_{\rm SO}}\right)^2 \delta_{\rm SO}^2 \frac{1}{ \tau_s} + D_e 
    (m^{*2} \epsilon_F \gamma_D)^2,
    \end{equation}
    where $1/\tau_{s} = 2 p_F^2  (\alpha_2^2 + (\alpha_1 - m^* \gamma_D \epsilon_F/2)^2 ) \tau  $.
    We introduced the dimensionless factor,
    $\delta_{\rm SO} = (Q_R^2-Q_D^2)/Q_{\rm SO}^2$ with
      $Q_{\rm SO}^2 = Q_D^2 + Q_R^2$ where 
  $Q_D$  depends
  on     Dresselhaus   spin-orbit coupling, $Q_D =  m^* (2 \alpha_1  - m^* \epsilon_F \gamma ) $. 
     $Q_R$ depends on   Rashba coupling: 
        $Q_R = 2 m^* \alpha_2$.
    Thus, for negligible cubic Dresselhaus spin-orbit coupling 
     the   the spin relaxation 
       length  increases when decreasing the wire width   $W$ as, 
       \begin{equation} \label{LsW}
       L_s (W)= \sqrt{D_e \tau_s (W)}\sim \frac{L_{\rm SO}^2}{W}.
       \end{equation}
This can be understood as follows: \cite{falko,schaepers,kettemann} 
    In a wire whose width exceeds the spin precession length $L_{\rm SO}$,  the area an electron 
       covers by diffusion in  time $\tau_s$ is  $W L_s$. To achieve spin relaxation, 
        this area    should be equal to the corresponding 2D spin relaxation area 
        $L_{s}(2D)^2$, where 
         $L_s (2D)= L_{\rm SO}/(2 \pi)$.
          Thus, the smaller the wire width, the larger the spin relaxation length 
           becomes,  $L_s \sim 
        (L_{\rm SO})^2/W$ in agreement with Eq.\,(\ref{LsW}).
         For larger wire widths, the spin diffusion equation can be solved 
         as well, and one finds that the   spin relaxation rate
          does not increase monotonously to the 2D limiting value but shows oscillations on the 
           scale $L_{\rm SO}$, which can be understood in analogy to  Fabry-P\'erot resonances.\cite{kettemann}
             For pure linear Rashba coupling that behavior can be derived analytically,
             in the approximation of a homogeneous spin density in transverse direction,
              yielding a relaxation rate given by 
           \begin{equation}
          \frac{1}{ \tau_s} (W) =     \frac{D_e}{2} Q_{\rm SO}^2  \left(1-  \frac{ \sin (Q_{\rm SO} W)}{Q_{\rm SO} W}\right), 
           \end{equation}
           where  $Q_{{\rm SO}} = 2 \pi/L_{\rm SO}$. 
           Furthermore, taking into account the transverse modulation of the 
            spin density by performing an exact diagonalization of the 
             spin  diffusion equation with the transverse boundary conditions, 
             Eq.\,(\ref{spincurrentBC}), one finds
              for $W>L_{\rm SO} $ modes which  are localized at the boundaries
              and have a lower relaxation rate
               than the bulk modes.\cite{raimondi,wenk2010}
                For pure Rashba spin relaxation we find that there
                 is a  spin-helix solution located at the edge
                  whose relaxation rate  $1/\tau_s =  .31/\tau_{s0}$ is smaller than  the 
                    spin relaxation rate of bulk modes $1/\tau_s =  7/16 \tau_{s0}$.
\subsection{Weak Localization Corrections}
\begin{figure}[ht]
\begin{center}
\epsfig{file=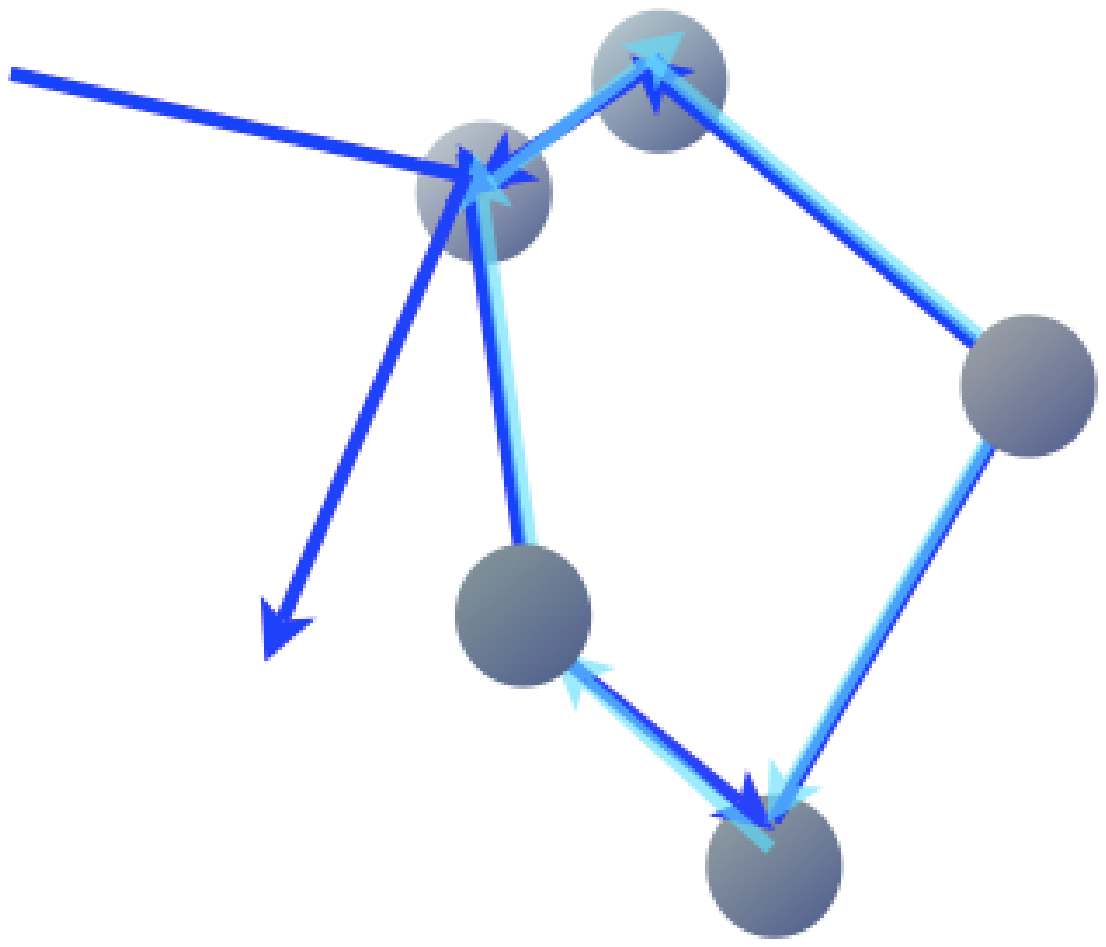,width=5cm}
\epsfig{file=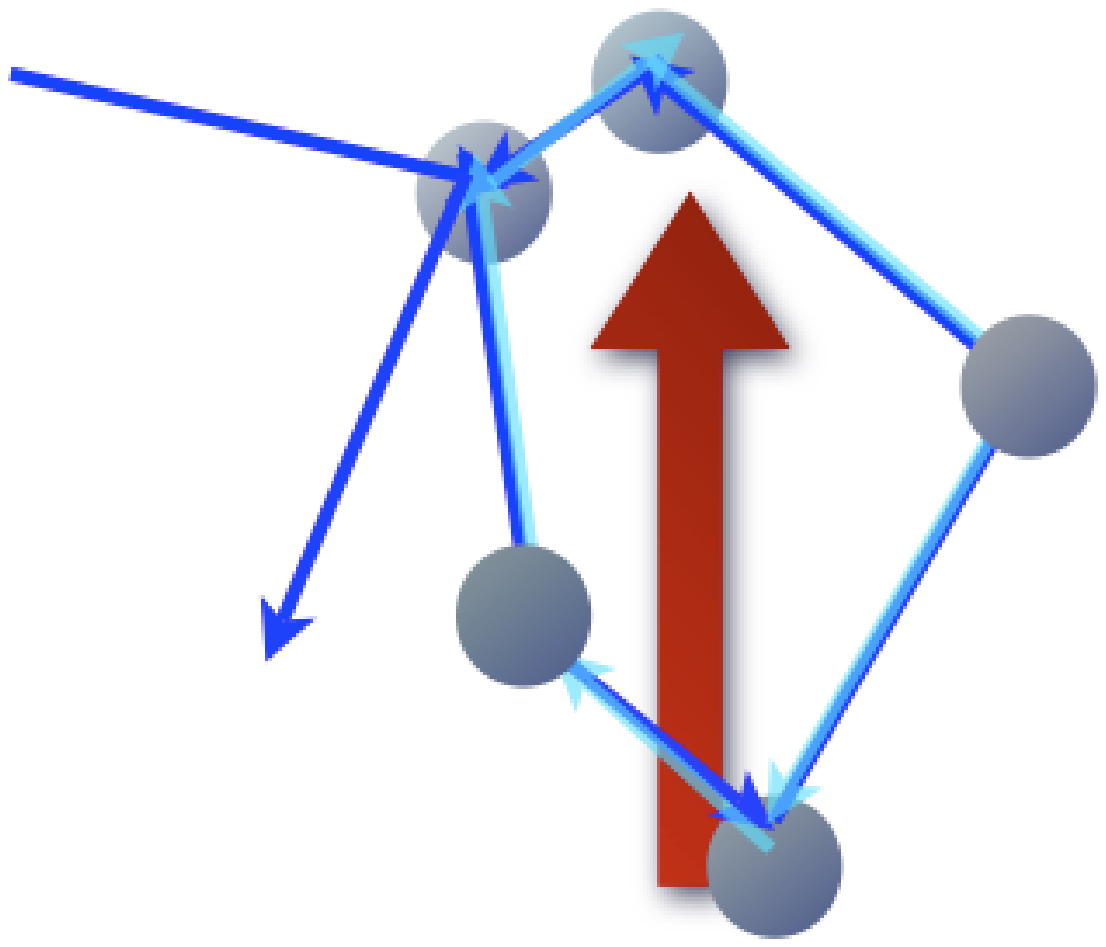,width=5cm}
\epsfig{file=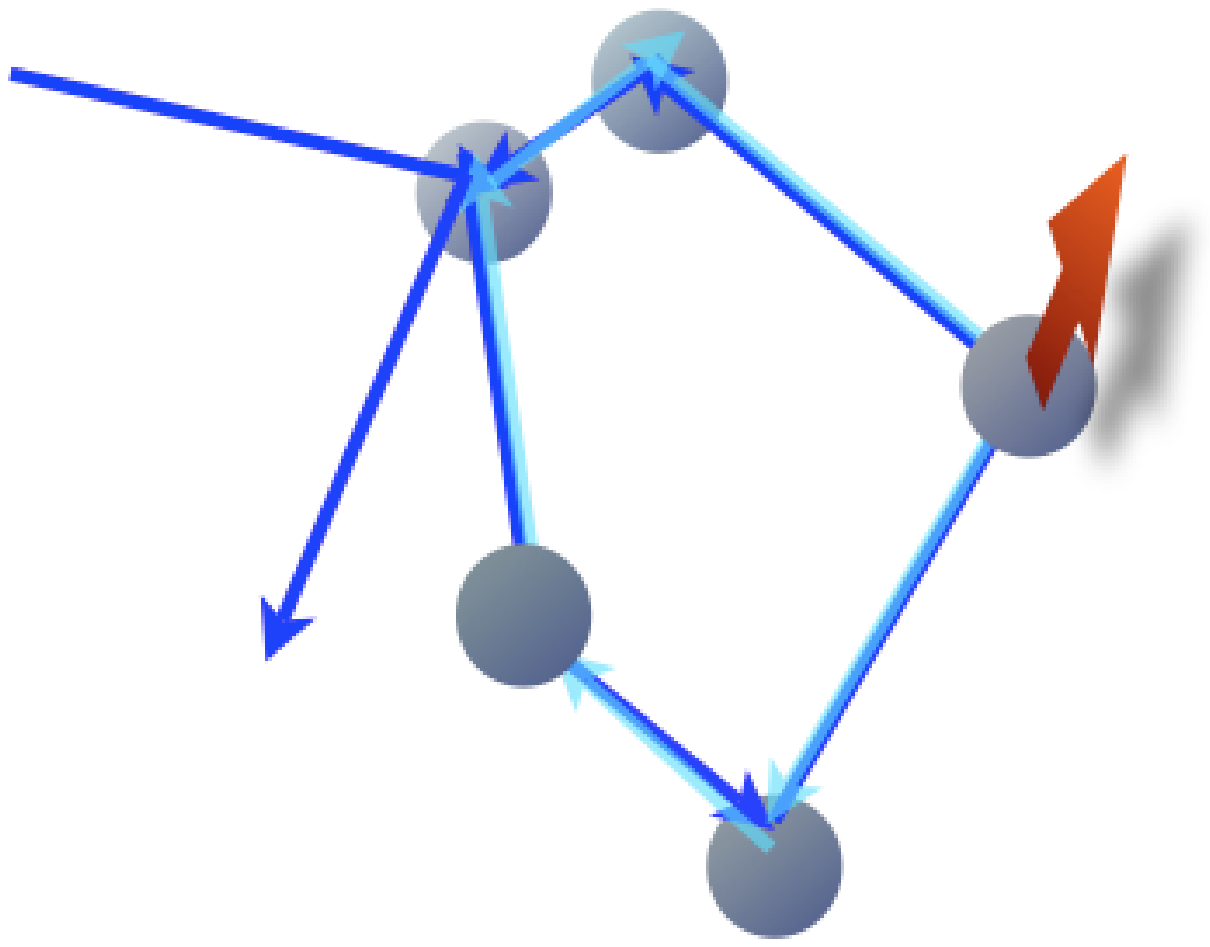,width=5cm}
 \caption{Electrons  can diffuse on closed paths,
   orbit clockwise or anticlockwise as indicated by the  light and dark blue arrows, respectively.
    Middle figure: Closed electron paths enclose a  magnetic flux from an external magnetic field, indicated as the red arrow, breaking time reversal symmetry. Right figure:  The scattering from a magnetic impurity spin, breaks the time reversal
        symmetry between the clock- and anticlockwise electron paths.} \label{Figwl}
\end{center}
\end{figure}

    Quantum interference of electrons in low-dimensional, disordered conductors
    results in corrections to the  electrical 
  conductivity $\Delta \sigma$.
     This quantum correction, the    weak localization 
   effect, is known to be 
    a very sensitive tool to study 
     dephasing and symmetry breaking mechanisms  in conductors.\cite{review1,review2,review3}
  The entanglement of spin and charge by   spin-orbit interaction
      reverses the effect of weak localization and 
      thereby enhances the conductivity, the weak antilocalization effect.
       The quantum correction to the conductivity $\Delta \sigma$ arises from the fact, that the
   quantum return probability to a given point ${\bf x}_0$ 
      after a time $t$, $P(t)$,   differs from 
    the classical return probability,
due to      quantum interference. 
  As the electrons scatter from impurities, 
   there is a finite  probability that they
    diffuse on closed paths, which  does increase 
  the lower the dimension of the conductor. Since an electron can move on such a closed
   orbit clockwise or anticlockwise as shown in light and dark blue in Fig.\,\ref{Figwl},  with equal probability, the probability amplitudes of
    both paths add coherently, if their length is smaller than the 
     dephasing length $L_{\varphi}$.
In a magnetic field, indicated by the
      red arrow in the middle Fig.\,\ref{Figwl}, the electrons acquire a  magnetic
       flux phase. This phase  depends on the direction in which the electron moves on the 
        closed path. Thus, the quantum interference is
         diminished in an external magnetic field since the area of closed paths  and thereby the flux
          phases  are randomly distributed
          in a disordered wire, even though  the  magnetic field can be constant.  
Similarly, the scattering from magnetic impurities breaks the time reversal invariance
between the two directions in which the closed path can be transversed. Therefore
 magnetic impurities diminish
the quantum corrections in proportion to the rate with which the electron spins
 scatter from them due to the exchange interaction, $1/\tau_{Ms}$, Eq.\,(\ref{tausm}).\\

\begin{figure}[ht]
\begin{center}
\epsfig{file=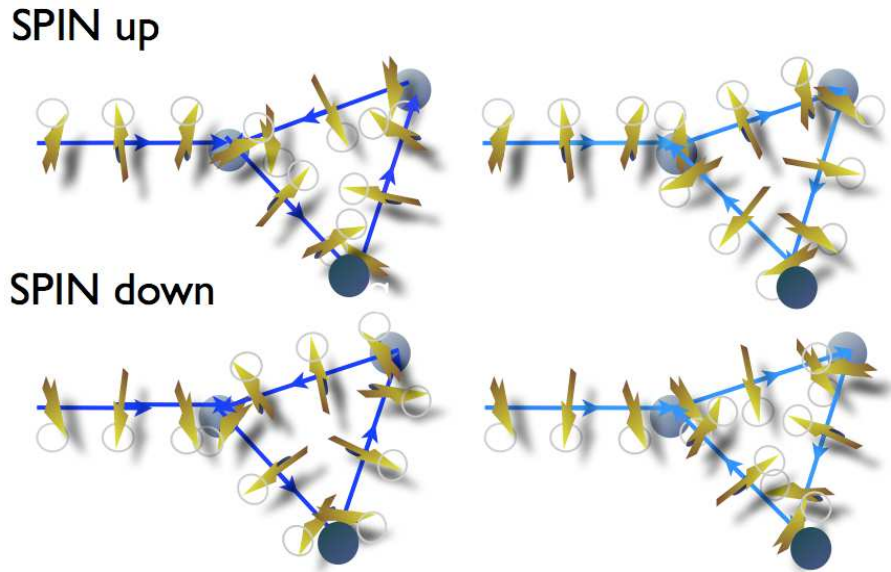,width=12cm}
 \caption{ As electrons diffuse, their spin
  precesses around the spin-orbit field, which changes its orientation,
   when the electron is scattered.  Electrons which enter  closed paths
    with  the same spin leave
    it therefore with a different spin  if they
    choose the path in the opposite sense, as  indicated by the light and dark blue arrows.
    However,  electrons which enter the closed path with opposite spin, and
      move through the closed path in  opposite sense, attain the same
       quantum phase.  This is a consequence of  time reversal invariance.} \label{Figawl}
\end{center}
\end{figure}

\begin{figure}[ht]
\begin{center}
\epsfig{file=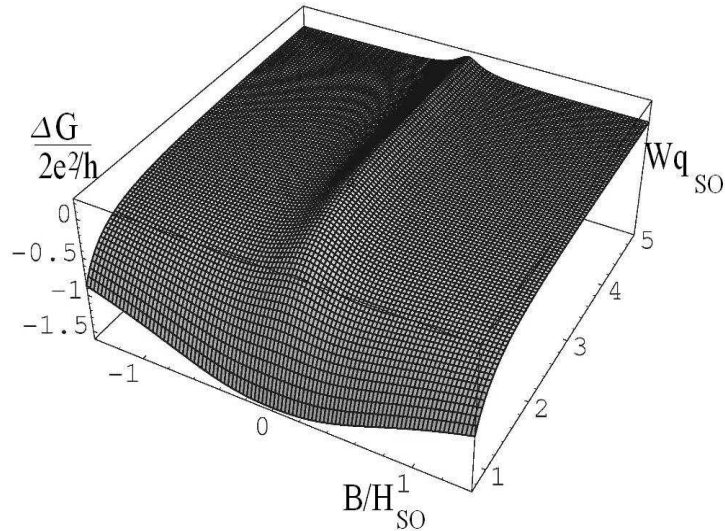,width= 10cm}
 \caption{
 The quantum conductivity correction  in units of $2 e^2/h$ as function of magnetic
 field $B$ (scaled with  bulk relaxation field $H_{s}$), and the wire width $W$ (scaled with
  $L_{\rm SO}/2 \pi$),  for pure Rashba coupling, $\delta_{\rm SO}=1$.} \label{Fig5}
\end{center}
\end{figure}

Thus, the quantum correction to the conductivity, $\Delta \sigma$
     is proportional to 
     the  integral over  all times
      smaller than  the dephasing time $\tau_{\varphi}$ of the quantum mechanical return probability $P (t) = \lambda_F^d \rho({\bf x},t)$, where 
       $d$ is  dimension of diffusion, 
         and $\rho$ is the electron density. 
        In the presence of spin-orbit scattering, the sign
         of the quantum correction  changes to  weak antilocalization
  as was as  predicted  by Hikami, Larkin, and Nagaoka \cite{nagaoka}
  for conductors with  impurities of heavy elements.  As   conduction 
    electrons scatter from such impurities, 
     the   spin-orbit interaction 
     randomizes their spin,  Fig.\,\ref{Figawl}.
 The resulting   spin relaxation suppresses   interference  of  time reversed paths  in  spin 
  triplet configurations, 
  while  interference in  singlet  configuration
    remains unaffected as indicated in Fig.\,\ref{Figawl}. Since  singlet interference
       reduces the electron's
      return probability it  enhances the conductivity, 
       the weak antilocalization effect. 
  Weak magnetic fields  suppress  also these singlet 
   contributions, reducing    the conductivity  and 	resulting in negative magnetoconductivity.          
  If the   host lattice of the electrons provides 
  spin-orbit interaction, the spin relaxation of DP or EY type does have the same effect of 
   diminishing the quantum corrections in the triplet configuration.
When the dephasing length $L_{\varphi}$ is smaller than the wire width $W$,
  the   quantum   corrections are determined by the interference of 2-dimensional 
  closed diffusion paths, and as a result,  the conductivity 
   increases logarithmically with $L_{\varphi}$ which increases itself as the temperature is lowered.
  At 
   low temperatures,
  the  electron-electron scattering is the dominating mechanism of spin dephasing,
   yielding  $L_{\varphi} \sim T^{-1/2}$.
  One can derive the magnetic field dependence of that quantum correction
   nonperturbatively.\cite{nagaoka,knap,miller,af01,lg98,golub} An approximate expression showing the
   logarithmic dependence explicitly is given by
   \begin{equation}  \label{wl2D}
 \Delta \sigma = - \frac{1}{2 \pi}\ln \frac{B+ \frac{4}{3} H_{Ms} +H_{\varphi}}{H_{\tau}  }
 +  \frac{1}{2 \pi}\ln \frac{B+H_{\varphi}+  H_{s}  + \frac{2}{3} H_{Ms} }{H_{\tau} }  
 +   \frac{1}{\pi} \ln  \frac{B+H_{\varphi} +  c H_{s} + \frac{2}{3} H_{Ms} }{ H_{\tau} },
 \end{equation}
 in units of $e^2/h$.
 All parameters are rescaled to dimensions of magnetic fields: 
 $H_{\varphi} = 1/(4 e D_e \tau_{\varphi}) = 1/(4 e L_{\varphi}^2)$,  $H_{\tau} = \hbar/(4 e D_e \tau)$,
  the spin relaxation field due to spin orbit relaxation, 
  $ H_{s} = \hbar/(4 e D_e \tau_{s})$,\cite{knap} and
   the spin relaxation field due to magnetic impurities  $ H_{Ms} = \hbar/(4 e D_e \tau_{Ms})$.
   Here $1/\tau_s$ is the DP relaxation rate in the 2D limit derived in the previous section.
     \cite{knap,iordanskii} The prefactor $c$ depends on the
      particular spin-orbit interaction. For linear Rashba-coupling, $c= 7/16$. Note that
     $7/16 \tau_s$ is the smallest spin relaxation rate of an inhomogeneous spin density distribution\cite{wenk2010} as derived in the section \ref{spindiffusion}.
     $1/\tau_{Ms}$ is the magnetic scattering rate from magnetic 
      impurities, Eq.\,(\ref{tausm}).
      Indeed we see that the first term does not depend on the 
     DP  spin relaxation rate. This term originates from 
              the interference of time reversed paths,  indicated in Fig.\,\ref{Figawl},
              which contributes to the quantum conductance in  the singlet
               state, 
       $\ket{S=0;m=0} = ( \ket{\uparrow \downarrow} - \ket{\downarrow \uparrow})/\sqrt{2}  
       $, the minus sign in front of the second term is the origin of the 
        change in sign in the weak localization correction. 
The other three terms are suppressed by 
 the spin relaxation rate, since they originate from interference in 
       triplet states, $\ket{S=1;m=0} = ( \ket{\uparrow \downarrow} + \ket{\downarrow \uparrow})/\sqrt{2}  ,\ket{S=1;m=1},\ket{S=1;m=-1}$ which do not conserve the spin symmetry.
        Thus, at strong spin-orbit induced spin  relaxation the last three terms are suppressed and the 
         sign of the quantum correction switches to weak antilocalization. 
In quasi-1-dimensional quantum wires which are coherent in transverse direction,
$W < L_{\varphi}$ the weak localization correction is further enhanced,
and increases linearly with the dephasing length $L_{\varphi}$.
 Thus, for  $W Q_{\rm SO} \ll 1$   the weak localization correction
    is  \cite{kettemann}
    \begin{align} \label{wl1D}
 \Delta \sigma ={}&   \frac{\sqrt{H_W}}{\sqrt{H_{\varphi}+\frac{1}{4} B^*(W) + \frac{2}{3} H_{Ms}}} - \frac{\sqrt{H_W} }{\sqrt{ H_{\varphi}
 + \frac{1}{4} B^*(W) +  H_{s}(W)  + \frac{2}{3} H_{Ms} }} 
 \nonumber \\   
 {}&- 2 \frac{\sqrt{H_W}}{ \sqrt{H_{\varphi}+ \frac{1}{4}B^*(W) + \frac{1}{2} H_{s}(W) + \frac{4}{3} H_{Ms} } },
 \end{align}
  in units of $e^2/h$.
  We defined $H_W = \hbar/(4 e W^2)$, and  the effective external magnetic field,
  \begin{equation} \label{beff}
   B^*(W) =  \left(1-1/\left(1+\frac{W^2}{3 l_B^2}\right)\right) B.
   \end{equation}   
    The spin relaxation  field $H_{s}(W)$ is  for $W < L_{\rm SO}$,
\begin{equation} \label{hso0}
 H_{s} (W) = \frac{1}{12}\left(\frac{W}{L_{\rm SO}}\right)^2 \delta_{\rm SO}^2 H_{s},
\end{equation}
    suppressed in proportion to 
      $(W/L_{\rm SO})^2$. Taking one transverse mode into account the quantum conductivity correction is plotted in Fig.\,\ref{Fig5} for different wire withs for pure Rashba SOC, showing the crossover from weak localization (positive magnetoconductivity) to weak antilocalization (negative magnetoconductivity).
In analogy to the  effective magnetic field, Eq.\,(\ref{beff}), the spin orbit coupling acts in quantum wires like  an effective  magnetic vector potential.\cite{falko} One  can expect that in  ballistic wires,
 $l_e>W$,  the spin relaxation rate is suppressed  
 in analogy  to the flux cancellation effect, which yields the 
weaker   rate, $1/ \tau_{s} = (W/C l_e) (D_e W^2/ 12 L_{\rm SO}^4)$,
 where $C= 10.8$.\cite{km02_1,km02_2,km02_3}
 A dimensional crossover from weak antilocalization to weak localization and a
   reduction of  spin relaxation  has recently been observed experimentally in quantum wires as we will 
    review in the next Section.

\section{Experimental Results on  Spin Relaxation Rate  in Semiconductor Quantum Wires}\label{sec:ExpResults}
\subsection{Optical Measurements}       
Optical time-resolved Faraday rotation (TRFR) spectroscopy \cite{schueller} has been used  to probe the  spin dynamics in an array of n-doped InGaAs wires by Holleitner et al. in Ref.\,[\onlinecite{holleitner,holleitnerjnp}]. The wires were dry etched  from  a quantum well grown in the [001]-direction  with a distance of 1~$\mu$m between the wires. Spin aligned charge carriers were created by absorption of circularly-polarized light. For normal incidence, the spins point then  perpendicular to the quantum well plane, in the growth direction [001]. The time evolution of the spin polarization was then measured with a linearly
          polarized pulse, see inset of  Fig.\,1c of Ref.\,[\onlinecite{holleitner}]. The
          time dependence  fits well with an exponential decay
           $\sim \exp (-\Delta t/\tau_s)$.     As seen in Fig.\,2a of Ref.\,[\onlinecite{holleitner}],
the thus measured lifetime  $\tau_s$  at fixed temperature $T = 5$ K of the spin polarization is enhanced when the   wire
 width $W$ is reduced\cite{holleitner}: While for $W >15$~$\mu$m it is $\tau_s = (12 \pm 1)$~ps,
  it increases for channels grown along the [100]- direction to almost $ \tau_s = 30$~ps, 
   and in the [110]- direction to about $\tau_s = 20$~ps.   Thus,  the experimental results show that 
      the  spin relaxation depends  on the patterning direction of the 
      wires: wires aligned along [100] and [010] show
equivalent spin relaxation times, which are generally longer than
the spin relaxation times of wires patterned along [110] and
[$\overline{1}$10].
The dimensional reduction could be seen already
for wire widths  as wide as $10$~$\mu$m, 
which is much wider than both the Fermi wave length and the elastic mean free path $l_e$ 
in the wires. This agrees well with the predicted reduction of the DP scattering rate, 
Eq.\,(\ref{SRW}) for wire widths smaller than the spin precession length $L_{\rm SO}$.
From the measured 2D spin diffusion length $L_s (2D) = (0.9 - 1.1)$~$\mu$m,  and its relation to the spin precession length Eq.\,(\ref{lso}), $L_{\rm SO} = 2 \pi L_s (2D)$,  we expect the crossover  to occur on a scale of $L_{\rm SO} = (5.7-6.9)$~$\mu$m as observed in Fig.\,2a of Ref.\,[\onlinecite{holleitner}].
From $L_{\rm SO} = \pi/m^* \alpha$ we get with $ m^* =0.064 m_e$ a spin-orbit coupling 
$\alpha = (5-6)$~meV\AA.
According to 
$L_s = \sqrt{D_e \tau_s}$, the spin relaxation length increases by a factor of 
$\sqrt{30/12} = 1.6$ in the [100]-, and by 
$\sqrt{20/12} = 1.3$ in the [110]- direction.\\
The spin relaxation time has been found to attain a maximum, however,  at about 
$W = 1\text{ $\mu$m} \approx L_s (2D)$, decaying appreciably for smaller widths.
	While a saturation of $\tau_s$ could be expected according to  Eq.\,(\ref{SRW})
	for diffusive wires,  due to cubic Dresselhaus-coupling, 
	a decrease is unexpected. Schwab et al., Ref.\,[\onlinecite{raimondi}], noted that
with      wire boundary conditions which do not conserve the spin of the conduction electrons
	one can obtain such a reduction. A mechanism for such spin-flip processes
	at the edges of the wire has not yet been identified, however.
 The magnetic field dependence of the 
 spin relaxation rate yields  further confirmation that the dominant 
  spin relaxation mechanism in these wires is  DPS: 
   It follows  the predicted behavior Eq.\,(\ref{DPSB}),
    as seen in Fig.\,3a of Ref.\,[\onlinecite{holleitner}], and the spin relaxation rate is
     enhanced to $\tau_s ( B= 1\text{ T}) = 100$~ps for all wire growth directions, at $T = 5$~K
      and wire widths of $W = 1.25$~$\mu$m.

  \subsection{Transport Measurements}
A dimensional crossover  from weak antilocalization to weak localization and a 
reduction of  spin relaxation  has recently been observed  experimentally in n-doped  InGaAs quantum wires,\cite{hu05,gh05}  in GaAs wires,\cite{lshh04} as well
as in AlGaN/GaN wires.\cite{lehnen}
The crossover indeed occurred in all experiments on the 
length scale of the spin precession length $L_{\rm SO}$.
We summarize in the following  the main results of these experiments.\\
Wirthmann et al., Ref.\,[\onlinecite{hu05}], measured the magnetoconductivity
of inversion-doped InAs quantum wells 
with a density of $n = 9.7 \times 10^{11}/\text{cm}^2$, and a measured effective mass of 
$m^*=0.04 m_e$. In the wide wires the  magnetoconductivity showed a pronounced  
weak antilocalization peak, which   agreed well with the 2D theory, \cite{iordanskii,knap}
with  a spin-orbit-coupling parameter of $\alpha = 9.3$~meV\AA.
	They observed a diminishment of the antilocalization peak 
	which  occurred for wire widths
	$W < 0.6$~$\mu$m, at $T= 2$~K, indicating a dimensional reduction of the DP spin relaxation rate.\\
	Sch\"apers et al. observed in Ga$_x$In$_{1 - x}$As/InP quantum wires  a complete crossover from weak  antilocalization to weak localization for wire widths below $W = 500$~nm.
	Such a crossover has also been observed in GaAs-quantum wires by Dinter et al., Ref.\,[\onlinecite{lshh04}].\\
	Very recently, Kunihashi et al., Ref.\,[\onlinecite{kunihashi}] observed the crossover from
	weak antilocalization  to weak localization in gate controlled  InGaAs quantum wires. 
	The asymmetric  potential normal to the quantum well 
	could be enhanced by application of a negative gate voltage, yielding an increase of the 
	SIA-coupling parameter $\alpha$, with decreasing carrier density, 
	as was obtained by fitting the magnetoconductivity of the quantum wells
	to the theory of 2D weak localization corrections of  Iordanskii et al., Ref.\,[\onlinecite{iordanskii}].
	Thereby, the  spin relaxation length $L_s= L_{\rm SO}/2 \pi$
	was found to decrease from $0.5$~$\mu$m to $0.15$~$\mu$m, which according to 
	$L_{\rm SO} = \pi/m^* \alpha$ corresponds to an increase of 
		$\alpha$ from $(20 \pm 1)$~meV\AA\enspace at electron concentrations of  $n = 1.4 \times 10^{12}/\text{cm}^2$
		to $\alpha = (60 \pm 1)$~meV\AA\enspace at electron concentrations of  $n = 0.3 \times 10^{12}/\text{cm}^2$.
	The magnetoconductivity of a sample with 95 
	quantum wires in parallel showed a clear crossover from weak antilocalization to 
	localization. 
	Fitting the data to Eq.\,(\ref{wl1D}) a corresponding decrease of the spin relaxation rate
	was obtained, which was observable already at large widths of the order
	of the spin precession length $L_{\rm SO}$ in agreement with the theory Eq.\,(\ref{SRW}).
	However, a saturation as obtained theoretically in  diffusive wires,
	due to cubic BIA-coupling was not observed. This might be due to the limitation of 
	Eq.\,(\ref{SRW}), to diffusive wire widths, $l_e < W$, while in ballistic wires
		a suppression also of the  spin relaxation due to cubic BIA-coupling can be expected, 
		since it vanishes identically in 1-D wires, see section \ref{1D}. 
		Also, an increase of the spin scattering rate in  narrower
		wires, $W < L_s(2D)$, was not  observed in contrast to the 
		results of the optical experiments, Ref.\,[\onlinecite{holleitner}],
		reviewed above.\\
The dimensional crossover has also been observed in the heterostructures
	of the wide gap semiconductor  GaN.\cite{lehnen} 
	The magnetoconductivity of 160 AlGaN/GaN- quantum wires were measured.
	The effective mass is $m^* =0.22 m_e$,  all wires were diffusive with $l_e < W$.
For electron densities of $ n \approx 5 \times 10^{12}/\text{cm}^2$ an increase from 
$L_s(2D) \approx 550$~nm to $L_s ( W  \approx 130\text{ nm}) > 1.8$~$\mu$m, 
and for densities $ n \approx 2 \times 10^{12}/\text{cm}^2$ an increase from
$L_s(2D) \approx 500$~nm to $L_s ( W  \approx 120\text{ nm}) > 1.$~$\mu$m was observed.
Using $L_s(2D) = 1/2 m^*\alpha$,  one obtains 
for both densities $n$, the spin-orbit coupling $ \alpha \approx 5.8$~meV\AA.
	A saturation of the spin relaxation rate could not be observed, 
	suggesting that  the cubic BIA-coupling is negligible  in these structures.\\
We note, that an enhancement of the spin relaxation rate  as in the optical experiments of narrow
	InGaAs quantum wires, Ref.\,[\onlinecite{holleitner}], was not observed in these
	AlGaN/GaN-wires. 
\section{Critical Discussion and Future Perspective}
The fact that  optical and transport measurements seem to find opposite behavior, enhancement and   suppression  of the spin relaxation rate, respectively, in narrow wires, calls for an  extension of  the theory to describe the crossover to ballistic quantum wires. This can be done, using the kinetic equation approach to the spin-diffusion equation,\cite{raimondi} a semiclassical approach, \cite{richter1,richter2} or an extension of the diagrammatic approach.\cite{wenk2010} In particular, the dimensional crossover of DPS due to cubic Dresselhaus coupling, which we found not to be suppressed in diffusive wires, needs to be studied for ballistic wires, $l_e > W$, as many of the experimentally  studied quantum wires are in this regime. Furthermore, using the spin  diffusion equation, one can study the dependence on the growth direction of quantum wires, and find more information on the magnitude of the various spin-orbit coupling parameters, $\alpha_1, \alpha_2, \gamma_D$, by comparison with the directional dependence found in both the optical measurements\cite{holleitner} of the spin relaxation rate, as well as in recent gate controlled transport experiments.\cite{kunihashi}\\
In narrow wires, corrections due to electron-electron interaction can become more important and influence especially the temperature dependence. Ref.\,[\onlinecite{holleitnerjnp}] reports a strong temperature dependence of the spin relaxation rate in narrow quantum wires. As shown in Ref.\,[\onlinecite{punnoose}],  the spin relaxation rates obtained from the spin diffusion equation and the quantum corrections to the magnetoconductivity can  be different, when corrections due to electron-electron interaction become important. As the DPS becomes suppressed in quantum wires other spin relaxation mechanisms like the EYS may become dominant, since it is expected that the dimensional dependence of EYS is less strong. In more narrow wires, disorder can also result in Anderson localization. Similar as  in quantum dots, \cite{khaetskii,khaetskiinuclearprl} this can yield  enhanced spin relaxation due to hyperfine coupling, Eq.\,(\ref{hyperfine}). The spin relaxation in metal wires is believed to be dominated by the EYS mechanism, which is not expected to show such a strong wire width dependence, although this needs to be explored in more detail. Even dilute concentrations of  magnetic impurities of less than 1 ppm, do yield measurable spin relaxation rates in metals and allow the study of the Kondo effect with unprecedented accuracy.\cite{kondo1,kondo2}
\section{Summary}
The spin dynamics and spin relaxation  of itinerant  electrons in disordered  quantum wires with spin-orbit coupling  is governed by the spin diffusion equation Eq.\,(\ref{diffusive}). We have shown that it can be derived by using classical  random walk arguments, in agreement with more elaborate derivations.\cite{raimondi,wenk2010} In semiconductor quantum wires all available experiments show that the motional narrowing mechanism of spin relaxation, the D'yakonov-Perel'-Spin relaxation (DPS) is the dominant mechanism in quantum wires whose width exceeds the spin precession length $L_{\rm SO}$. The solution of the spin diffusion equation reveals existence of persistent spin helix modes when the linear BIA- and the SIA-spin-orbit coupling are of equal magnitude. In quantum wires which are more narrow than the spin precession length $L_{\rm SO}$ there is an effective alignment of the spin-orbit fields  giving rise to  long living spin density modes  for arbitrary ratio of the  linear BIA- and the SIA-spin-orbit coupling. The resulting reduction in the spin relaxation rate results in a change in the sign of the  quantum corrections  to the conductivity. Recent experimental results  confirm the increase of the spin relaxation rate in wires whose width is smaller than $L_{\rm SO}$, both the direct optical measurement of the spin relaxation rate, as well as transport measurements. These show a dimensional crossover from weak antilocalization to weak localization as the wire width is reduced. Open problems remain, in particular in narrower, ballistic wires, were optical and transport measurements seem to find opposite behavior of the spin relaxation rate: enhancement,  suppression, respectively. The experimentally observed reduction of spin relaxation in quantum wires opens new perspectives for spintronic applications, since the spin-orbit coupling and therefore the spin precession length remains unaffected, allowing a better control of the itinerant electron spin. The observed directional dependence moreover can yield more detailed information about the spin-orbit coupling, enhancing the spin control for  future spintronic devices further.
\newpage
\section*{Symbols}
    \begin{description}
\item[$\tau_0$] elastic scattering time
\item[$\tau_{ee}$] scattering time due to electron-electron
interaction
\item[$\tau_{ep}$] scattering time due to electron-phonon
interaction
\item[$\tau$] total scattering time $1/\tau=1/\tau_0+1/\tau_{ee}+1/\tau_{ep}$.
\item[$\hat{\tau}_s$] spin relaxation tensor
\item[$D_e$] diffusion constant, $D_e=v_F^2\tau/d_D$, where $d_D$ is the dimension of diffusion.
\item[$l_e$] elastic mean free path
\item[$L_{{\rm SO}}$] spin precession length in 2D. The spin will be oriented again in the initial direction after it
moved ballistically the length $L_{{\rm SO}}$.
\item[$Q_{{\rm SO}}$] $=2\pi/L_{{\rm SO}}$
\item[$L_s$] spin relaxation length, $L_s(W)=\sqrt{D_e\tau_s(W)}$ with
$L_s(W) \mid_{W \to \infty}=L_s(2D) = L_{\rm SO}/2\pi$
\item[$L_\varphi$] dephasing length
\item[$\alpha_1$] linear (Bulk inversion Asymmetry (BIA) = Dresselhaus)-parameter
\item[$\alpha_2$] linear (Structural inversion Asymmetry (SIA) =Bychkov-Rashba)-parameter
\item[$\gamma_D$] cubic (Bulk inversion Asymmetry (BIA) = Dresselhaus)-parameter
\item[$\gamma_g$] gyromagnetic ratio
\end{description}

\begin{acknowledgments}
We thank V. L. Fal'ko, F. E. Meijer, E. Mucciolo, I. Aleiner, C. Marcus, T. Ohtsuki, K. Slevin,  J. Ohe, and A. Wirthmann for helpful  discussions. This research was supported by DFG-SFB508 B9 and by WCU ( World Class University ) program through the Korea Science and Engineering Foundation funded by the Ministry of Education, Science and Technology (Project No. R31-2008-000-10059-0).
\end{acknowledgments}

\bibliographystyle{apsrev}
\bibliography{WenkKettemannBook2010.bib}
\end{document}